\documentclass[twocolumn]{aastex631}

\usepackage{amsmath}
\usepackage{amssymb}	
\usepackage{CJK}
\usepackage{float}

\usepackage{color}

\newcommand{\rc}{r_\mathrm{c}}
\newcommand{\rt}{r_\mathrm{t}}

\begin{document}
\begin{CJK*}{UTF8}{gbsn}
\title{Pair-Instability Gap Black Holes in Population III Star Clusters: Pathways, Dynamics, and Gravitational Wave Implications}

\correspondingauthor{Long Wang}
\email{wanglong8@sysu.edu.cn}

\author{Weiwei Wu (吴维为)}
\affiliation{School of Physics and Astronomy, Sun Yat-sen University, Daxue Road, Zhuhai, 519082, China}

\author[0000-0001-8713-0366]{Long Wang (王龙)}
\affiliation{School of Physics and Astronomy, Sun Yat-sen University, Daxue Road, Zhuhai, 519082, China}
\affiliation{CSST Science Center for the Guangdong-Hong Kong-Macau Greater Bay Area, Zhuhai, 519082, China}

\author{Shuai Liu (刘帅)}
\affiliation{School of Electronic and Electrical Engineering, Zhaoqing University, Zhaoqing 526061, People's Republic of China}

\author{Yining Sun (孙逸宁)}
\affiliation{School of Physics and Astronomy, Sun Yat-sen University, Daxue Road, Zhuhai, 519082, China}

\author{Ataru Tanikawa}
\affiliation{Center for Information Science, Fukui Prefectural University, 4-1-1 Matsuoka Kenjojima, Eiheiji-cho, Fukui 910-1195, Japan}

\author{Michiko Fujii}
\affiliation{Department of Astronomy, Graduate School of Science, The University of Tokyo, 7-3-1 Hongo, Bunkyo-ku, Tokyo 113-0033, Japan}



\begin{abstract}

The detection of the gravitational wave (GW) event GW190521 raises questions about the formation of black holes within the pair-instability mass gap (PIBHs). We propose that Population III (Pop III) star clusters significantly contribute to events similar to GW190521. We perform $N$-body simulations and find that PIBHs can form from stellar collisions or binary black hole (BBH) mergers, with the latter accounting for 90\% of the contributions. Due to GW recoil during BBH mergers, approximately 10-50\%
of PIBHs formed via BBH mergers escape from clusters, depending on black hole spins and cluster escape velocities. The remaining PIBHs can participate in secondary and multiple BBH formation events, contributing to GW events. Assuming Pop III stars form in massive clusters (initially 100,000 $M_\odot$) with a top-heavy initial mass function, the average merger rates for GW events involving PIBHs with 0\% and 100\% primordial binaries are $0.005$ and $0.017$ \(\text{yr}^{-1} \text{Gpc}^{-3}\), respectively, with maximum values of $0.030$ and $0.106$ \(\text{yr}^{-1} \text{Gpc}^{-3}\). If Pop III stars form in low-mass clusters (initial mass of $1000M_\odot$ and $10000 M_\odot$), the merger rate is comparable with a 100\% primordial binary fraction but significantly lower without primordial binaries. 
We also calculate the characteristic strains of the GW events in our simulations and find that about 43.4\% (LISA) 97.8\% (Taiji) and 66.4\% (Tianqin) of these events could potentially be detected by space-borne detectors, including LISA, Taiji, and TianQin. The next-generation GW detectors such as DECIGO, ET, and CE can nearly cover all these signals.

\end{abstract}

\keywords{Population III stars(1285) --- N-body simulations(1083) --- Supernova remnants(1667) --- Gravitational wave astronomy(675)}


\section{Introduction} \label{sec:intro}

Studies of stellar evolution suggest that very massive stars with masses ranging between about 100-300 $M_\odot$ can undergo the pair-instability supernovae (PISNe) \citep{Fowler1964, Rakavy1967, Fraley1968, Ober1983, Bond1984, El1986, Fryer2001, Langer2012}.
In the 1960s, \cite{Fowler1964} proposed the instability of electron-positron pairs within the interiors of stars. When the helium core mass of a star falls within the range of about 45 to 135 $M_\odot$, high-energy photons in its electromagnetic radiation field can convert into electron-positron pairs \citep{Rakavy1967,Fraley1968}. This reduces the radiation pressure from the core, failing to counteract the inward gravitational force and causing core collapse. 
Stars with helium core masses between about 65 and 135 $M_\odot$ explode entirely due to the immense heat produced, leaving no remnants, and thus predict a pair-instability mass gap of black holes (PIBHs) \citep{Ober1983,Bond1984,El1986}. 
For stars with helium core masses between about 45 and 65 $M_\odot$, after the collapse induced by the instability of electron-positron pairs, further thermonuclear reactions release even more high-energy photons, causing radiation pressure to rise again. This repeated oscillation triggers a series of strong pulsations, effectively reducing the masses of helium and heavier elements. This phenomenon is known as pulsational pair-instability supernovae (PPSNe) \citep{Woosley2007,Woosley2017,Heger2002}.

Defining the limits of the pair-instability mass gap depends on various factors, such as different supernova explosion models, stellar metallicity, stellar winds, core evolutionary rates, binary separation evolution, and rotation. In some theoretical models and $N$-body simulations, this mass gap is estimated to lie between about 50 $M_\odot$ and 120 $M_\odot$ \citep{Belczynski2016,Spera2017,Spera2019}. 

However, the LIGO and Virgo collaborations detected a gravitational wave (GW) event, GW190521 \citep{Abbott2020,Abbott2020b,Abbott2022}, where the signal indicated the merger of a binary BHs (BBHs) with masses of approximately 85 and 66 $M_\odot$, resulting in an intermediate-mass black hole (IMBH) of around 142 $M_\odot$. 
At least one of the black holes (BHs) involved in the GW190521 event falls within the mass gap caused by PISNe (PPSNe). More GW candidates of PIBHs have been found in recently updated catalogs, including GW190403\_051519 and GW190426\_190642 in GWTC-2.1 \citep{Abbott2024}
, GW200220\_061928 and a few more with large uncertainties in GWTC-3 \citep{Abbott2023} and GW200129\_114245 in 4-OGC \citep{Nitz2023}. In this study, we define BHs with masses in the range of 65--120 $M_\odot$ as PIBHs.

Several scenarios have been proposed to explain the formation of PIBHs. 
\cite{mandel2022rates} and \cite{spera2022compact} provide comprehensive overviews for an in-depth exploration of the impact of different stellar physics assumptions on the formation of PIBHs their merger rates from different channels.
A common one suggests that PIBHs may form through the mergers of binary stars or BBHs in star clusters, including young, globular, and nuclear clusters. These scenarios have been explored in numerous studies \citep{Antonini2019,DiCarlo2019,DiCarlo2020b,DiCarlo2020a,DiCarlo2021,Gerosa2019,Banerjee2021,Banerjee2022,Kremer2020,Mapelli2021,Gerosa2021,Fragione2020,Liu2021,Gonzalez2021,Gonzalez2022,Rodriguez2019,Rodriguez2022,Gondan2021,Arca-Sedda2021,Costa2022,Rizzuto2022,Rose2022,Anagnostou2022,DallAmico2021,Torniamenti2022,Torniamenti2024,Ballone2023}.
If a massive star on a giant branch collides and merges with a star on the second main-sequence band, the result may be an evolutionary star with a super large hydrogen envelope and a He core mass below 45 $M_\odot$. In such a case, the star can avoid PISN and collapse into a PIBH.
In addition, PIBHs can form via mergers of BBHs, but may gain a high velocity due to GW recoil.
When the host star cluster has a sufficiently high escape velocity, such as nuclear star clusters, PIBHs can be retained in the system.
Then the stellar dynamics can bring the PIBHs and other BHs together and form a GW190521-like progenitor. 

Recent studies show that Population (Pop) III stars may also be important progenitors of PIBHs. 
Firstly, Pop III stars may directly evolve into the low-mass end of PIBHs.
The pair-instability mass gap can vary with the combination of H-rich envelope collapse and a low $^{12}C(\alpha,\gamma)^{16}O$ rate \citep{Farmer19, Costa2021}\footnote{\cite{Takahashi18} has first pointed out that the $^{12}C(\alpha,\gamma)^{16}O$ rate significantly affects helium core mass ranges within which PISNe happen.}, or with slight changes to the overshooting parameter and model \citep{Iorio2023}.
Pop III stars formed at high redshifts with extremely low metallicity $Z<10^{-8}$ have much weaker winds than the Population I/II stars and thus have relatively higher remnant masses for a certain zero-age main-sequence (ZAMS) mass.
With low efficient convective overshoot, Pop III stars with ZAMS masses of $60$--$85 M_\odot$ can keep their initial mass before the core collapse and thus can form PIBH with mass up to $85~M_\odot$. \citep{Costa2023,Santoliquido2023,Tanikawa2020,Tanikawa2021,Tanikawa2022,Tanikawa2024}.
In addition, the Pop III stars may have a top-heavy IMF \citep{Stacy2016,Chon2021,Chon2024,Latif2022}.
Therefore, the fraction of massive BH remnants for Pop III stars can be so high that the contribution of PIBHs can be significant.

Additionally, Pop III stars may form in star clusters and produce PIBHs via stellar and BBH mergers \citep{Sakurai2017,Wang2022,Liu2024,Mestichelli2024arXiv}. They may also form PIBHs via binary stellar evolution and dynamical hardening in nuclear star clusters \citep{LiuBoyuan2020,LiuBoyuan2021,LiuBoyuan2024}. Due to their zero metallicity, Pop III stars are typically more massive and more likely to form binary systems, which may evolve into massive BBH systems. \cite{kinugawa2021formation} simulations estimated the event rate of such BBH mergers as 0.13--0.66 \(\text{ yr}^{-1} \text{Gpc}^{-3}\). The detection rate of the coalescing Pop III BBHs is 140(68) events \(\text{yr}^{-1} \) for the flat (Salpeter) initial mass function (IMF), and the GW signals produced by the BBH mergers from Pop III stars are detectable by current and next-generation GW detectors \citep{kinugawa2014possible}. 

Pop III stars are not only interesting for GW detection, but also important for understanding the star formation at the early Universe. 
It is also possible to detect massive Pop III stars and their PISN signals via electromagnetic observation.
Recently, a possible PISN from a massive Pop III star was detected \citep{Xing2023}, indicating that a very massive Pop III star can form.
With James Webb Space Telescope, it is possible to detect massive Pop III stars at high redshifts \citep{Larkin2023,Bovill2024,Venditti2024,Zackrisson2024}.

This work focuses on the formation of PIBHs from Pop III star clusters through a series of $N$-body simulations, aiming to understand the detailed formation scenarios of PIBHs and the properties of GW events with PIBHs originating from these clusters. 
It is part of a series of studies following \cite{Wang2022} and \cite{Liu2024}.
In their work, Pop III star clusters are placed inside a mini dark matter halo, following the zoom-in cosmology model from \cite{Sakurai2017}.
\cite{Wang2022} carried out a series of models by varying the lower limit of the IMF, primordial mass segregation, the density of the mini dark-matter halo, and the central concentration of the density profile of clusters.
They found that hierarchical mergers of massive stars can occur to form very massive stars, and they eventually evolve to IMBHs in the center of Pop III star clusters.
This occurs when the initial central density is sufficiently high, and this criterion is independent of the other parameters.
\cite{Liu2024} continued to investigate the mergers of BBHs with IMBHs inside triples in Pop III star clusters, with additional 10 times more simulations with and without primordial binaries.
They found that it is possible to detect IMBH-BH and PIBH-BH mergers by future space-borne detectors such as TianQin \citep{luo2016tianqin}, LISA \citep{amaro2017laser}, Taiji \citep{ruan2020taiji} and new generation of ground-based detectors such as Einstein GW detector \citep{punturo2010einstein}.

While these works focused on IMBH formation and thus investigated massive Pop III star clusters with an initial mass of $10^5 M_\odot$, which is like the upper end, 
most Pop III star clusters have a total mass between $10$--$1000 M_\odot$ as shown in the cosmological simulation for Pop III star formation \citep{Hirano2015}, 
These low-mass clusters may not contribute to IMBH formation but can be important for the formation of PIBH.
Therefore, in this work, we analyze the PIBH formation and corresponding GW mergers by analyzing the simulation data of \cite{Liu2024}, representing the massive end of Pop III star clusters, and also perform new simulations with less massive clusters, which are expected to be the major population of Pop III star clusters. 
We also consider the effect of GW recoils and spins of BHs, which is important for GW events with PIBHs.

In Section~\ref{sec:method}, we introduce the numerical method for performing the $N$-body simulations of Pop III star clusters. Section~\ref{sec:result} shows the details of the results, including the formation mechanism, mass distribution, and formation time of the PIBHs, GW events with the influence of GW kick PIBH, and the detection ability of different GW detectors. In Section~\ref{sec:discussion}, we discuss how our results compare with other channels. Section~\ref{sec:conclusion} draws the conclusion.

\section{Method} \label{sec:method}

\subsection{$N$-body method} \label{subsec:Petar}

Using the same numerical methods as in \cite{Wang2022} and \cite{Liu2024}, we simulated the long-term evolution of Pop III star clusters with the high-performance $N$-body code \textsc{petar} \citep{Wang2020b}. 
This code employs a hybrid integrator with three algorithms adapted for different ranges of separation of objects. 
For long- and medium-range interactions, it uses the particle–tree and particle–particle algorithms \citep{Oshino2011} and the Framework for Developing Particle Simulator \citep{Iwasawa2016,Iwasawa2020}. 
For short-range interactions, such as close encounters and orbital evolution of multiple systems, it implements the slow-down algorithmic regularization method and the \textsc{sdar} code \citep{Wang2020a}. 

Additionally, the code incorporates single and binary stellar evolution packages based on \textsc{sse}/\textsc{bse} \citep{Hurley2000,Hurley2002} and its extended versions, including \textsc{bseemp} \citep{Tanikawa2020}, to evolve the mass, stellar types, and other parameters of individual stars and binaries. 
In our work, we use the \textsc{bseemp} code 
to model the single and binary stellar evolution of Pop III stars. 
\textsc{bseemp} offers fitting formulas for the evolution tracks of massive stars in extremely metal-poor environments. 
Therefore, we could trace the stellar wind mass loss, stellar-type changes, BH formation, and binary interaction in Pop III stars clusters with the minimum metallicity $Z = 2 \times 10^{-10}$.
The metallicity of Pop III stars in this study is set to be this minimum.

The code also supports the use of \textsc{galpy} \citep{Bovy2015} to include external tidal forces from dark matter potentials. 
We use it to generate the mini dark matter potential hosting Pop III star clusters.

\subsection{Initial condition}
\label{sec:Initial_condition}

We established the initial conditions for the Pop III star clusters based on the models from \cite{Liu2024}. 
In addition to the models with initial total mass \( M = 10^5 M_\odot \), we performed new simulations for \( M = 10^3 \) and \( 10^4 M_\odot \), keeping all other parameters the same. 
The following outlines the initial conditions of our model sets.

For all three masses, we adopt an initial half-mass radius of $r_h = 1$~pc, which is typical for present-day star clusters. 
We use the \cite{michie1963}-\cite{king1966} model for the density profile of star clusters with a high central concentration by setting $W_0 = 9$, where $W_0$ is a dimensionless measure of system concentration defined as the ratio of the core radius ($\rc$) and the tidal radius ($\rt$). 
\cite{Wang2022} found that this concentration parameter can lead to the formation of IMBHs, consistent with the findings of \citep{Sakurai2017}. 

We adopt the IMF of Pop III star clusters as a top-heavy function with a single power-law profile and an power index of approximately -1, following the investigation of hydrodynamical models for Pop III star formation from \cite{Stacy2016,Chon2021,Chon2024,Latif2022}.
Considering the uncertainties regarding the mass range of the IMF for Pop III stars, we assume that the minimum and maximum mass limits are $1$ and $150~M_\odot$ in this study.
This IMF would lead to a significantly larger proportion of massive stars.
Eventually, BHs dominate the mass of the cluster and drive the clusters to ``dark clusters'' \citep{Banerjee2011}. 

Although the upper mass limit of Pop III star in our simulations is $150~M_\odot$, stars with masses on the order of \( 10^3 M_\odot \) can form via hierarichal mergers during the early evolution of Pop III star clusters with a total mass of $10^5 M_\odot$, as discussed in detail in \cite{Wang2022}. 
Therefore, our work also accounts $10^3 M_\odot$ Pop III stars. 

For extremely massive Pop III stars with masses on the order of \(10^4 M_\odot \), as predicted by the extreme value statistics in \cite{Chantavat2023}, their formation is rare. Most Pop III star clusters have a total mass between $10$--$1000 M_\odot$ based on the cosmological simulations of Pop III star formation in \cite{Hirano2015}. If $10^4 M_\odot$ stars do form, they would most likely exist as single stars rather than within star clusters, due to the limited total progenitor gas mass and the low star formation efficiency of Pop III stars. These cases fall outside the scope of our simulations.

For the dark matter halo structure, we adopted the NFW profile \citep{NFW1996}, where the virial mass $M_{\text{vir}} = 4 \times 10^7 \, M_{\odot}$ and the virial radius $ r_{\text{vir}} = 280 \, \text{pc} $,  which follows the model A from \cite{Sakurai2017}. 
We assumed that the star clusters formed at redshift $z = 20 $.

For $M = 10^3$ and $10^4 M_\odot$, we carry out two subsets of simulations with and without primordial binaries, labeled as `bf1' and `bf0', respectively.
There are no constraints on the properties of primordial binaries of Pop III stars. Therefore, we assume their binary properties are similar to those of young metal-rich stars and adopt the distributions of initial periods, mass ratios, and eccentricities based on \citep{Sana2012}. The initial binary fraction is 100 percent.

For each combination of $M$ and the primordial binary fractions, we carry out approximately 300 individual simulations with different randomly sampled initial masses, positions, and velocities. The summary of the initial condition is shown in Table~\ref{tab:init}. 

\begin{table*}
	\caption{ The initial condition of models of Pop III clusters. }
\label{tab:init}
	\begin{tabular}{lcccccc} 

        \hline
		Model name & m100000-bf1 & m100000-bf0 & m10000-bf1 & m10000-bf0 & m1000-bf1 & m1000-bf0 \\
  		\hline Initial Mass [$M_{\odot}$ ]
		 & \multicolumn{2}{c}{$1 \times 10^{5}$} & \multicolumn{2}{c}{$1 \times 10^{4}$} & \multicolumn{2}{c}{$1 \times 10^{3}$} \\
		
		Primordial binary fraction & 1 & 0 & 1 & 0 & 1 & 0 \\
		number of models & 354 & 287 & 282 & 300 & 300 & 300 \\

		\hline
	\end{tabular}
\end{table*}

\subsection{Pop III stellar evolution model}

We breifly introduce our Pop III stellar evolution model, which is summarized in Appendix A of \cite{Tanikawa2022}. \cite{Tanikawa2020} constructed fitting formulas of stellar evolution tracks based on 1D hydrodynamical simulations performed using the \textsc{hoshi} code \citep{2016MNRAS.456.1320T,2018ApJ...857..111T,Takahashi19,Yoshida19}. They developed two types of fitting formula: the L and M models. The L model assumes larger convective overshooting parameters compared to the M model. 

In the L model, Pop III stars with masses of $8-10M_\odot$ and $50-160M_\odot$ evolve into red supergiant stars, while those with masses of $10-50M_\odot$ end as blue supergiant stars. This behavior is similar to the Pop III star evolution model proposed by \cite{Marigo2001}. In contrast, the M model predicts that Pop III stars with masses of $8-160M_\odot$ end as blue supergiant stars, which is similar to the Pop III stellar evolution model described by \cite{2021MNRAS.502L..40F}. The reason for this difference is as follows. In the L model, more efficient convective overshooting allows more fresh hydrogen to be supplied to the convective core during the main-sequence phase. As a result, when the star ends its main-sequence phase, it has a more massive He core. This larger He core leads to higher luminocity during the post main-sequence phase, causing the star to expand to a larger radius.

The key difference between the L and M models is that isolated binaries cannot form PIBH-BH mergers in the L model, while they can in the M model \citep{Tanikawa2021}. 
In the L model, a $\sim100~M_\odot$ star, which could potentially leave behind a PIBH, expands to $\sim3000~R_\odot$ during its post-main-sequence phase. At this stage, it interacts with its companion star through common envelope evolution and/or stable mass transfer, ultimately losing its hydrogen envelope. As a result, the star leaves behind only a $\lesssim 40~M_\odot$ BH, rather than a PIBH. If the star does not interact with its companion due to a large orbital separation, it can evolve into a PIBH. However, in such cases, the resulting BBH cannot merge within the Hubble time because of their wide separation. Consequently, in the L model, PIBH-BH mergers can only form through dynamical interactions. 

In contrast, in the M model, a $\sim100~M_\odot$ Pop III star expands to only a few $10~R_\odot$ during its post-main-sequence phase. This limited expansion prevents interaction with its companion star, even if their separation is $\sim100~R_\odot$. As a result, the star and its companion do not undergo common envelope evolution or stable mass transfer, allowing the star to collapse into a PIBH. Because the separation between the two stars remains small, the resulting BBH can merge within the Hubble time. 
Thus, isolated binaries in the M model can form merging PIBH-BH systems, unlike those in the L model.

In this paper, we only choose the L model, as our goal is to investigate how  efficiently dynamical interactions can form PIBH-BH mergers.

\subsection{BH spin model}
\label{sec:spin}

BH spins can largely increase GW recoil kick velocities which BBH merger remnants receive at their merger moments. However, \textsc{petar} has not yet extracted BH spins, although \textsc{bse} and \textsc{bseemp} internally evolve stellar spin angular momenta and their resulting BH spins. In order to estimate GW recoil kicks, we assign BH spins to BBH members in a post-processing way, referring to BH spins generated through isolated binary evolution calculated with \textsc{bseemp} \citep{Tanikawa2021, Tanikawa2022}. In this section, we overview how to calculate BH spins in \textsc{bseemp}, and how to assign BH spins to BBH members in our simulaiton.

\textsc{bseemp} adopts the same framework of stellar spin evolution as \textsc{bse}. Stellar spins decrease by mass loss of stellar winds, common envelope evolution, and supernova, increase by mass accretion from their companion stars, and change through tidal interactions. Detail models related to BH spins are different between \textsc{bse} and \textsc{bseemp}. Here, we describe these models in \textsc{bseemp}. Stellar wind models in \textsc{bseemp} are similar to those of \textsc{bse}. They are based on \cite{2010ApJ...714.1217B}. The detail implementation of \textsc{bseemp} is described in \cite{Tanikawa2021}. The implementation of common envelope evolution is the same between \textsc{bse} and \textsc{bseemp}. When a star loses its envelope, its spin is assumed to be zero. Similarly to \textsc{bse}, the supernova model in \textsc{bseemp} refers to \cite{2012ApJ...749...91F} with modifications of PISN and PPISN \citep{2020A&A...636A.104B}. When BH prognitors lose their masses at their core collapses, they also lose their spin angular momenta, depending on the mass loss as described in \cite{Tanikawa2022}. The prescription of stable mass transfer in \textsc{bseemp} is the same as in \textsc{bse}. For tidal interaction, \textsc{bseemp} adopts the same as in \textsc{bse} for convective damping, while it chooses a fitting formula proposed by \cite{2010ApJ...725..940Y} and \cite{2018A&A...616A..28Q} for radiative damping.

We simulate isolated binary evolutions by means of \textsc{bseemp}, and tabulate spin parameters of BBH members as a function of semi-major axes, eccentricitis, primary BH masses, and secondary BH masses at the formation time of these BBHs. Referring to this BH spin function, we assign a spin parameter to each BH of a BBH at the moment when this BBH is formed.

This BH spin model is justified to some degree by the following simple consideration. According to \cite{Tanikawa2021, Tanikawa2022}, tidal interations most affect BH spins in \textsc{bseemp}. A BH progenitor is spun-up by tidal interactions, and collapses to a BH, taking into its spin angular momentum due to little mass loss. Such tidal interactions strongly depend on the semi-major axis of its BBH progenitor; the tidal interactions become larger as the semi-major axis become smaller. The semi-major axis of the BBH progenitor is similar to that of the resulting BBH, since a BH prongeitor loses little mass at its core collapse. Thus, a BH spin depends on the semi-major axis of its BBH at the formation time. A BH spin depends on the eccentricity, primary BH mass, and secondary BH mass of its BBH in a minor way. That's the reason why the BH spin function include these parameters as its arguments.

To estimate the spins of BHs for each BBH merger in our star cluster simulations during post-processing, we adopt the following approach. For each BBH in the simulation, we identify a BBH from isolated binary evolution models (produced with \textsc{bseemp}) that has the most similar orbital parameters at the time of BBH formation, including semi-major axis, eccentricity, and BH masses. The spin values from the matched reference BBH are then assigned to the corresponding BBH in our simulation.

To efficiently find the closest matching binary, we first construct a four-dimensional KD tree using weighted values of these four parameters for the set of reference binaries. The parameter weights are determined as follows:  
We initially build an unweighted KD tree, use it to find the closest matches in the reference set for each target binary, and compute the average difference for each parameter across all matches. These average differences represent the typical matching error for each parameter. The weights are then set to the inverse of these differences, so that parameters with larger typical errors contribute less during matching. 
The normalized weights for the four parameters are 0.0112 and 0.0088 for the two BH masses, 0.0022 for the semi-major axis, and 0.9777 for the eccentricity.  
The eccentricity receives the highest weight because it varies within a narrow range, resulting in a smaller matching error compared to the other parameters.
Finally, we construct a weighted KD tree with these optimized weights and use it to identify the closest reference binary for each target BBH, from which we assign the spin values.

This method is reasonable for the following reasons:
For BBHs formed from non-interacting binaries (e.g. little tidal interaction between the two components) and dynamically formed BBHs, the spins of the BHs are identical to those of BHs formed through single stellar evolution. In this case, there is no difference in spins between the star cluster environment and the isolated binary environment.

For BBHs formed from interacting binaries, the spins of the BHs are determined by the binary evolution history. In star clusters, binaries can experience dynamical interactions with surrounding stars, particularly in systems like triples. These interactions can alter the orbital parameters of the binaries, leading to differences in the orbital parameter distributions of BBHs formed in star clusters compared to those formed in isolated environments.
However, once a binary becomes tight and enters the interaction phase, the influence of distant perturbations on its orbital parameters becomes negligible. If a strong perturbation occurs at close range, it may cause the binary to merge before forming a BBH or to widen, preventing the formation of a BBH capable of merging via GW radiation.
Therefore, for BBHs in star cluster simulations that originate from interacting binaries and are capable of merging, we expect that most of them evolve similarly to isolated binaries once they enter the interaction phase. As a result, we can reliably estimate their spins by referencing BBHs from isolated binary evolution models with matching orbital parameters (semi-major axis, eccentricity, and component masses) at the time of BBH formation.

In this way, while we may incorrectly assign spins to BHs from strongly perturbed interacting binaries, such systems represent only a small fraction of the total population. Thus, the statistical influence of these errors on spin-related effects, such as GW recoil kicks, remains acceptable.

\section{Result}
\label{sec:result}

\subsection{Formation channels of PIBHs} 

\begin{figure*}
    \includegraphics[width=0.33\textwidth]{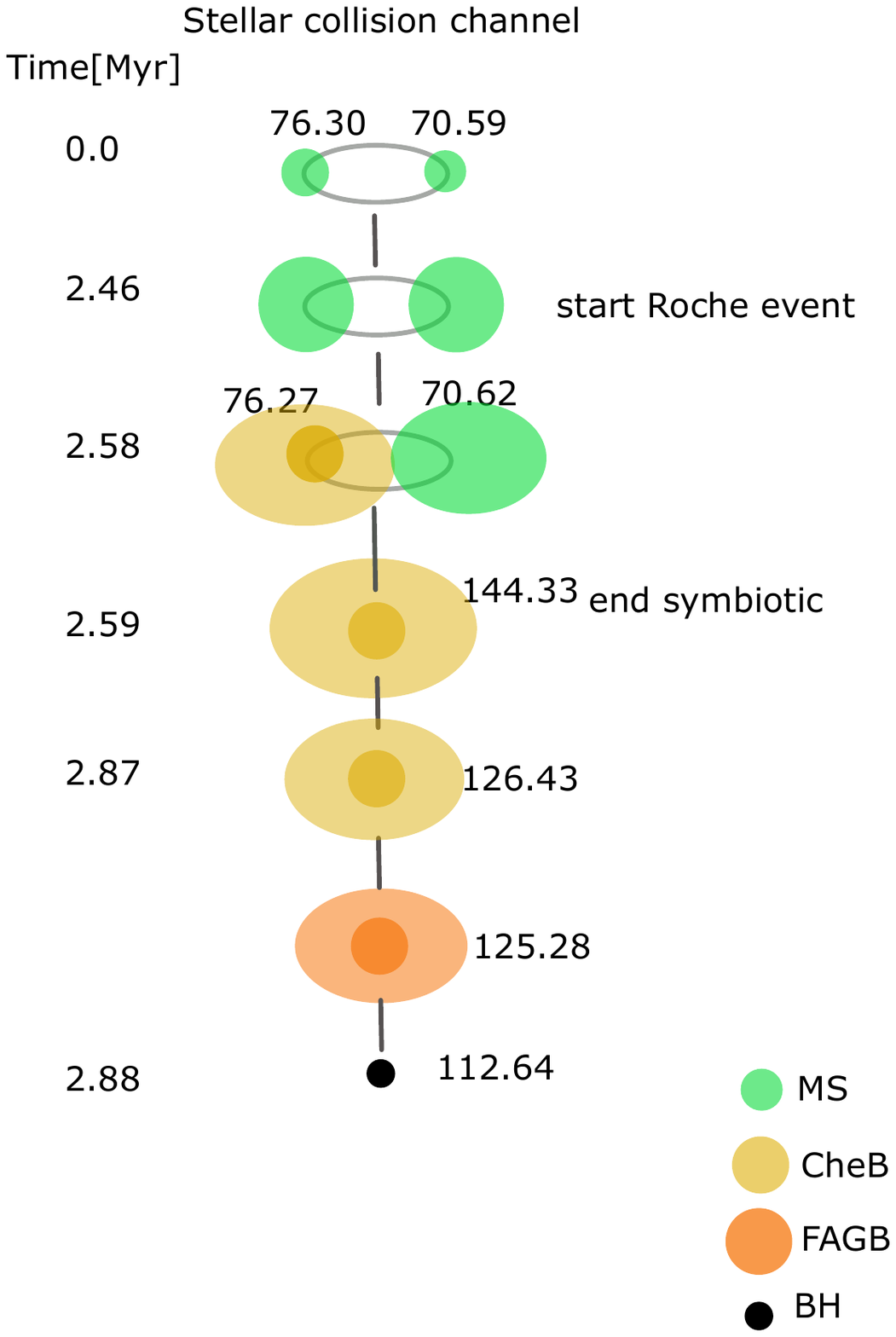} 
    \includegraphics[width=0.33\textwidth]{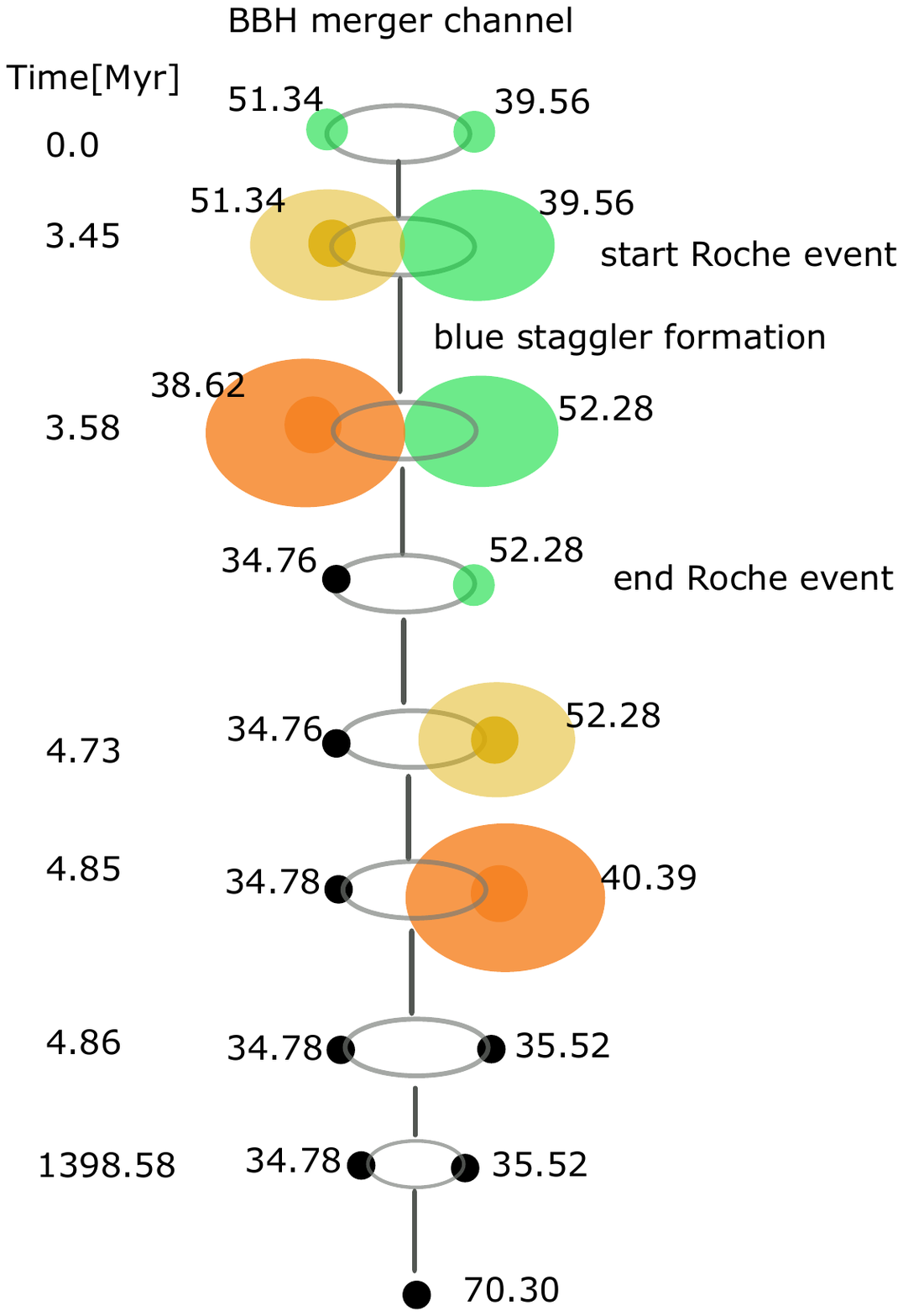}
    \includegraphics[width=0.33\textwidth]{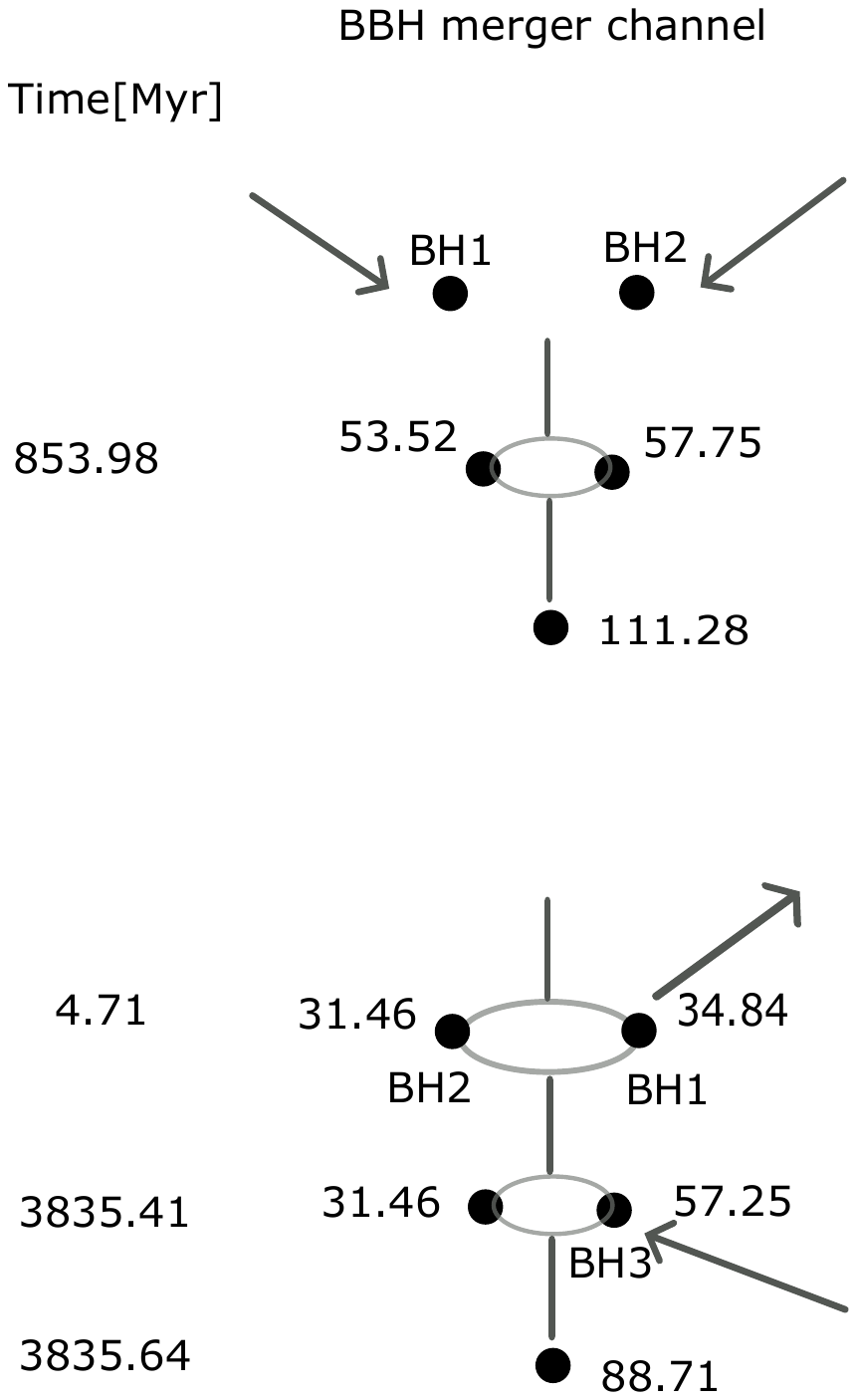}\\
    \caption{The evolution histories of three PIBH formation channels are illustrated, with example events taken from the m100000-bf1 models. In the left panel, a PIBH forms via a stellar collision, evolving into a PIBH without experiencing PISN. The middle panel shows a BBH formed through binary stellar evolution that merges into a PIBH via GW radiation. In the right panel, a dynamically formed BBH merges into a PIBH, with potential exchanges of BH components occurring during the process. } 
    \label{fig:pibh-formation}
\end{figure*}

We first analyzed the formation channels of PIBHs in Pop III star cluster models, which can occur through two mechanisms: stellar collisions and BBH mergers. 
Figure~\ref{fig:pibh-formation} illustrates these channels. 
The left panel depicts an example of PIBH formation via binary stellar evolution and collision, taken from the model m100000-bf1. 
In this scenario, a binary system with two main-sequence stars, masses \( 76.3 M_\odot \) and \( 70.59 M_\odot \), undergoes a Roche lobe overflow event, merging into a core helium-burning (CHeB) star with a total mass of \( 144.33 M_\odot \). 
This star evolves into an Asymptotic Giant Branch (AGB) star, reducing its mass to \( 125.28 M_\odot \). 
Compared to a similar-mass AGB star formed from a single star, this merged star has an oversized hydrogen envelope and a core mass below \( 45 M_\odot \), allowing it to avoid PISN and evolve into a PIBH. 
Ultimately, the star undergoes gravitational collapse, forming a BH of \( 112.64 M_\odot \). This formation scenario has also been observed in young star clusters \citep{Spera2019,DiCarlo2019,DiCarlo2020b,Kremer2020}.

In contrast, the middle and right panels illustrate PIBH formation via BBH mergers, which occur through two sub-channels. 
The first case, shown in the middle panel, involves a BBH formed through binary stellar evolution. 
Initially, a binary system with two main-sequence stars, with masses \( 51.34 M_\odot \) and \( 39.56 M_\odot \), undergoes a Roche lobe overflow event. 
One star begins CHeB, evolves into an AGB star, and eventually collapses into a BH with a mass of \( 34.76 M_\odot \). 
The other star follows a similar evolutionary path, becoming a BH with a mass of \( 35.52 M_\odot \). 
The close separation of the BBH triggers GW radiation, leading to a merger that forms a PIBH with a mass of \( 70.30 M_\odot \).

The second case, illustrated in the right panel, shows PIBH formation via dynamically formed BBHs. 
If the BBHs are sufficiently close, they can merge into a PIBH. 
In some instances, a third BH can exchange with one component of the BBH. 
In the example shown in the plot, a BH with a mass of \( 57.25 M_\odot \) replaces one of the BHs with a mass of \( 34.84 M_\odot \). 
Ultimately, the BBH merges to form a PIBH with a mass of \( 88.71 M_\odot \).

Table~\ref{tab:number} shows PIBH formation numbers across models. With primordial binaries, BBH mergers produce 10-30 times more PIBHs than stellar collisions, while PIBH formation through stellar collisions is zero without primordial binaries. Additionally, BBH mergers produce far fewer PIBHs in models without primordial binaries, indicating that BBH mergers in primordial binary models are primarily driven by primordial binaries.

With primordial binaries,  PIBHs formed via BBH mergers scale linearly with $M$, decreasing by a factor of 10 for every 10-fold decrease in cluster mass.
Without primordial binaries, PIBH formed via BBH mergers shows weaker dependence on $M$, suggesting that dynamically formed BBH mergers are less influenced by cluster mass.

\begin{table*}
    \centering
    \caption{The first two rows show the average numbers of PIBHs formed via the stellar collision and BBH merger channels per cluster. The next two rows compare the subsequent PIBH-BH merger events for the two formation channels. In the BBH merger channel, PIBHs with GW kick velocities exceeding the cluster escape velocity ($v_{\text{esc}}$) are excluded in PIBH-BH merger event counts. The values of $v_{\text{esc}}$ and the ejected fraction of PIBHs ($f_\mathrm{e}$), with and without considering BH spins, are also provided. The last row shows all BBH merger numbers per cluster. }
    \label{tab:number}
    \begin{tabular}{lcccccc} 

        \hline
	Model name & m100000-bf1 & m100000-bf0 & m10000-bf1 & m10000-bf0 & m1000-bf1 & m1000-bf0 \\
  	\hline 
	PIBH formed via stellar collision & 1.89 & 0 & 0.43 & 0 & 0.02 & 0 \\
  	PIBH formed via BBH mergers & 49.09 & 0.43 & 4.60 & 0.23 & 0.56 & 0.08 \\
        PIBH-BH merger (stellar collision) & 0.08 & 0 & 0.02 & 0 & 0 & 0 \\
        PIBH-BH merger (BBH mergers) & 0.548 & 0.181 & 0.043 & 0.033 & 0.003 & 0 \\
    $v_{\text{esc}}$ [km/s] & 65 & 67 & 57 & 60 & 56 & 56 \\
    $f_{\mathrm{e}}$ without spin [\%] & 3.7 & 21.3 & 6.3 & 15.4 & 12.2 & 0 \\
    $f_{\mathrm{e}}$ with spin [\%] & 9.9 & 45.9 & 12.7 & 27.7 & 22.0 & 20 \\
    All BBH merger & 96.78 & 4.96 & 9.49 & 0.27 & 1.05 & 0.02 \\
    
	\hline
    \end{tabular}
\end{table*}

\subsection{Mass distribution}

\begin{figure}
	\includegraphics[width=\columnwidth]{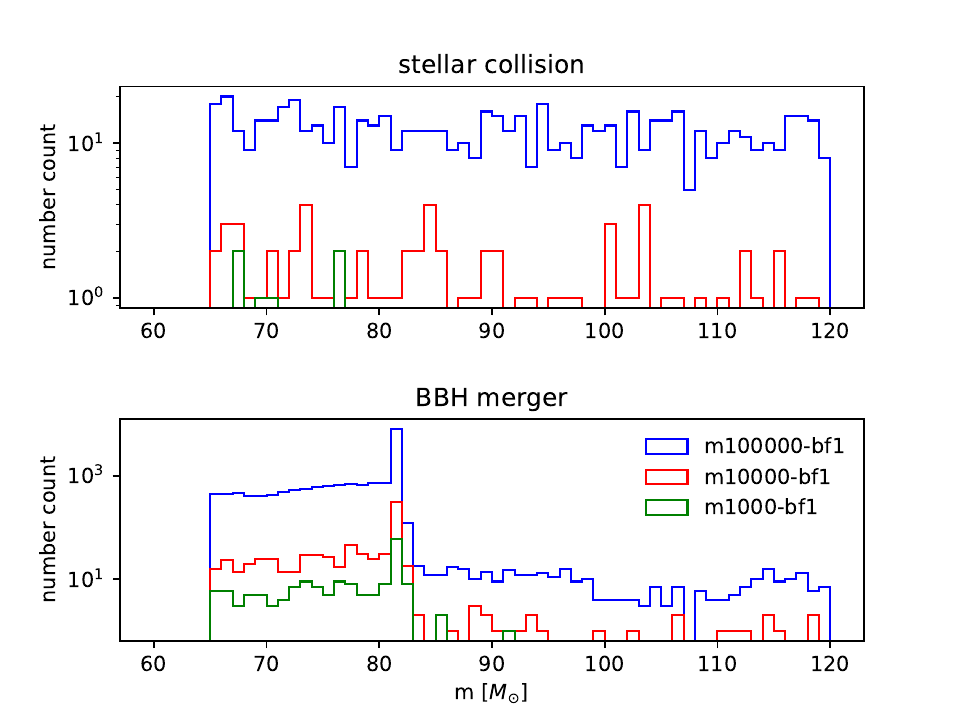}
    \caption{Mass spectrum of PIBHs from simulations with three different $M$ and with primordial binaries. The upper and lowe panels show PIBHs formed through stellar collision and BBH merger channels, respectively.}
    \label{fig:mass distri_1}
\end{figure}

We investigate the mass distribution of PIBHs formed through the stellar collision and BBH merger channels. 
Figure~\ref{fig:mass distri_1} shows the mass spectrum of PIBHs formed through these two channels, respectively. 
The data are from all simulations with primordial binaries.
The BBH merger channel contributes a significantly larger number of PIBHs than the stellar collision channel. 

The BH mass distribution formed through the stellar collision channel is relatively uniform, while there is a significant peak around 80 $M_\odot$ in the BBH merger channel. 

The observed peak is attributed to a specific aspect of our model: BHs that form directly from zero-age main-sequence stars exceeding a certain mass threshold, approximately 80 $M_\odot$, have a consistent mass of about 40.5 $M_\odot$. These BHs constitute a significant portion of the overall BH population. 
The mass spectrum has a similar shape across different $M$, showing no dependeonce on cluster mass.

It is important to note that while BHs with a mass of 40.5 $M_\odot$ represents the dominate BH population, the maximum mass of BHs formed through the single stellar evolution of Pop III stars in our simulations is not $40.5 M_\odot$.
BHs that form directly from zero-age main-sequence stars with initial masses exceeding approximately 80 $M_\odot$, have the same mass of about 40.5 $M_\odot$. 
In contrast, stars with initial masses between 50-80 $M_\odot$ collapse to form BHs with masses ranging from $40-60$ $M_\odot$, as shown in Figure 1 of \cite{Wang2022}. 

Our definition of PIBHs refers to BHs that cannot form through single stellar evolution. 
Consequently, we define the lower limit of the mass gap starting at 65 $M_\odot$, which is slightly higher than 60 $M_\odot$. 
This value aligns with the definitions used in previous studies \citep[e.g.][]{Heger2003,Belczynski2016,Spera2017,Woosley2017,Woosley2019,Giacobbo2018}.

\begin{figure}
	\includegraphics[width=\columnwidth]{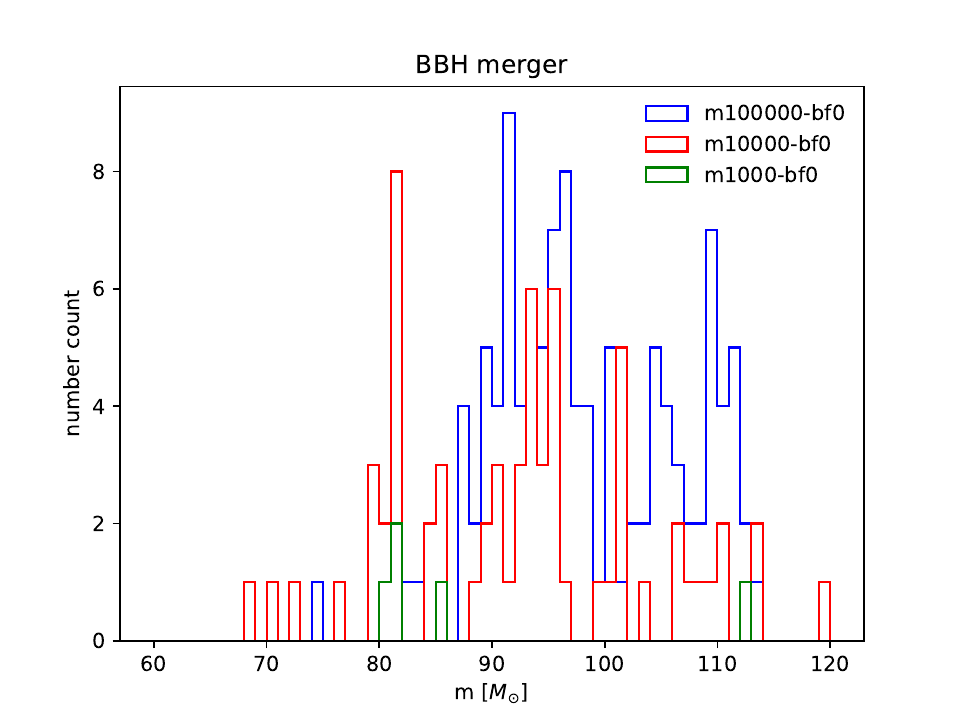}
    \caption{Similar to Figure~\ref{fig:mass distri_1}, but show models without primordial binary. In these models, PIBHs are generated only through the BBH merger channel. 
    }
    \label{fig:mass distri_0}
\end{figure}

Figure~\ref{fig:mass distri_0} presents the results of models without primordial binaries. In these clusters, no PIBHs formed through the stellar collision channel; all PIBHs were produced via the BBH merger channel. 
Comparing Figure~\ref{fig:mass distri_1} and Figure~\ref{fig:mass distri_0}, clusters with primordial binaries generate significantly more PIBHs, particularly low-mass PIBHs below about $80~M_\odot$. 
This suggests that low-mass PIBHs from BBHs with primordial binaries are primarily driven by binary stellar evolution. 
With the \cite{Sana2012} period distribution, some primordial binaries can evolve into BBH mergers through pure binary stellar evolution, without requiring dynamical encounters to shrink binary orbits in star clusters.
Therefore, models without primordial binaries lack this population.

Without primordial binaries, PIBHs form solely through dynamically formed BBHs, with their mass spectrum determined by individual BHs.
Since the BH mass spectrum is similar across different $M$, the PIBH mass spectrum shows no obvious difference.

\subsection{Formation time}

\begin{figure}
    \includegraphics[width=\columnwidth]{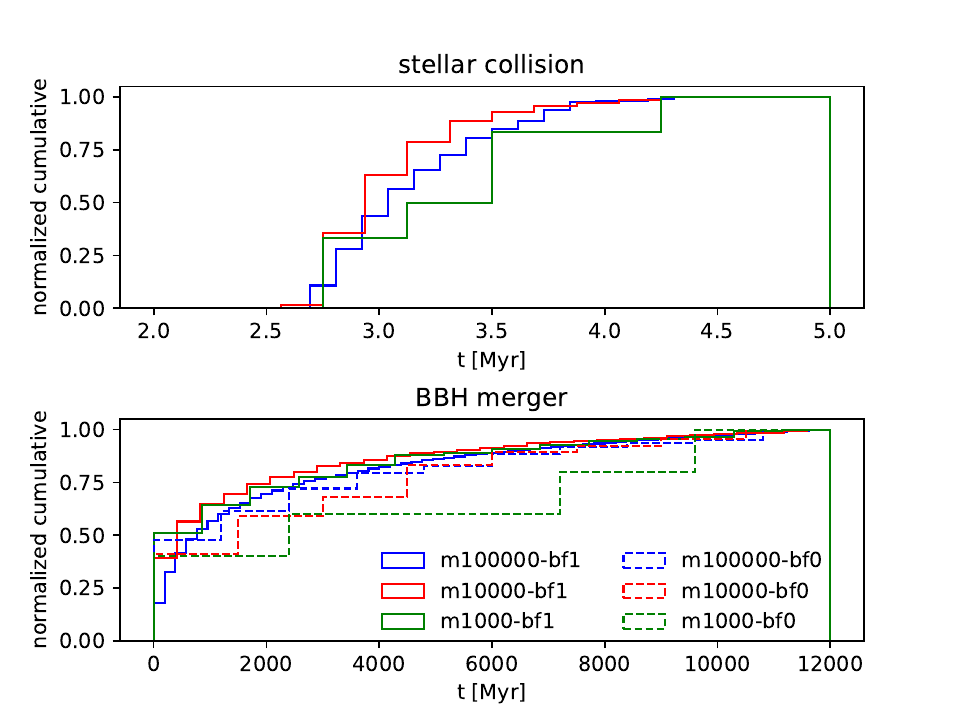}
    \caption{The normalized cumulative distribution of the formation time of PIBHs through two channels. }
    \label{fig:formation t_1}
\end{figure}

We also analyzed the formation times of PIBHs, comparing models with and without primordial binaries. Understanding the formation time of PIBHs is crucial, as it directly impacts their observability and the expected redshift range for potential detections. 
Figure~\ref{fig:formation t_1} shows the normalized cumulative distribution of the formation time of PIBHs for models with and without primordial binaries.
The formation of PIBHs predominantly occurs within the first 4 Myr through the stellar collision channel. 
In contrast, PIBHs formed via the BBH merger channel mainly merge after 5 Myr and continue forming until the end of simulations. 
From the lower panel, it is evident that the models including primordial binaries and those without them exhibit similar overall trends in the distribution of formation time of PIBHs. 
This indicates that the presence of primordial binaries does not significantly change the merger time distribution of BBHs. 
One possible explanation is that BBHs formed through dynamical processes have a timescale similar to primordial binaries, leading to comparable time distribution patterns in both models. 

\subsection{PIBH-BH merger}

To create events similar to GW190521, PIBHs must form a BBH with another BH (PIBH-BH) and merge through GW radiation. 
In this section, we explore these events in our models.

\subsubsection{The influence of gravitaitonal wave recoil}

If PIBHs form via the BBH merger channel, they experience GW recoil and may be ejected from star clusters, preventing the formation of new BBH mergers that could produce events like GW190521.
GW recoil, or GW kick, occurs when the newly formed BH receives a significant momentum kick due to asymmetric GW emission during the merger.
As two BHs orbit and merge, the emitted GWs are not always symmetrically distributed, creating a reactive force that imparts velocity to the resulting BH. 
This recoil can be substantial, significantly altering the BH's position from its original location.
The \textsc{petar} code version (master branch, similar to v1.0) used in our simulations lacks the GW kick effect, so we must analyze how this impacts events similar to GW190521.

The magnitude of GW kick velocity depends on several key factors: the mass ratio of the merging BHs, their spin magnitudes, and the orientation of their spins relative to the orbital angular momentum. We calculate the recoil velocity magnitude, \( v_{\mathrm{k}} \), using the formula from \citep{gerosa2016precession}:
\begin{equation}
v_{\mathrm{k}} = \sqrt{v_{\mathrm{m}}^2 + 2 v_{\mathrm{m}} v_{\perp} \cos \zeta + v_{\perp}^2 + v_{\parallel}^2}, 
\end{equation}
where \( v_{\mathrm{m}} \), \( v_{\perp} \), and \( v_{\parallel} \) represent the kick velocity components due to mass asymmetry in the orbital plane, spin asymmetry perpendicular to the orbital angular momentum, and the parallel kick velocity component, respectively. \( \zeta \) is the angle between \( v_{\mathrm{m}} \) and \( v_{\perp} \).
The formulas of the three velocity components are as follows:

\begin{equation}
\begin{aligned}
v_{\mathrm{m}} = &A \eta^2 \frac{(1 - q)(1 + B \eta)}{1 + q}, \\
v_{\perp} = &H \eta^2 \Delta_{\parallel}, \\
v_{\parallel} = &\; 16 \eta^2 \Bigg(\Delta_{\perp} \Big(V_{11} + 2 V_A \tilde{\chi}_{\parallel} + 4 V_B \tilde{\chi}_{\parallel}^2 + 8 V_C \tilde{\chi}_{\parallel}^3 \Big) \\
&\; + 2 \tilde{\chi}_{\perp} \Delta_{\parallel} \Big(C_2 + 2 C_3 \tilde{\chi}_{\parallel} \Big) \Bigg) \cos^2 \Theta,
\end{aligned}
\end{equation}

where \( q \) is the mass ratio of the smaller BH ($m_2$) to the larger one ($m_1$), and \( \tilde{\chi}_{\parallel} \) and \( \tilde{\chi}_{\perp} \) denote the parallel and perpendicular components of the spin, respectively. 
The coefficients are as follows: \( A = 1.2 \times 10^4 \, \text{km/s} \), \( B = -0.93 \) \citep{gonzalez2007maximum}, \( H = 6.9 \times 10^3 \, \text{km/s} \),  \(\zeta = 145^\circ\) \citep{lousto2008further}, \( V_{11} = 3677.76 \, \text{km/s} \), \( V_A = 2481.21 \, \text{km/s} \), \( V_B = 1792.45 \, \text{km/s} \), \( V_C = 1506.52 \, \text{km/s} \) \citep{lousto2012gravitational}, \( C_2 = 1140 \, \text{km/s} \) and \( C_3 = 2481 \, \text{km/s} \) \citep{lousto2013nonlinear}. 
These values are derived from numerical-relativity simulations of BH mergers and characterize various aspects of the recoil velocity generated during BBH mergers.

\begin{figure}
	\includegraphics[width=\columnwidth]{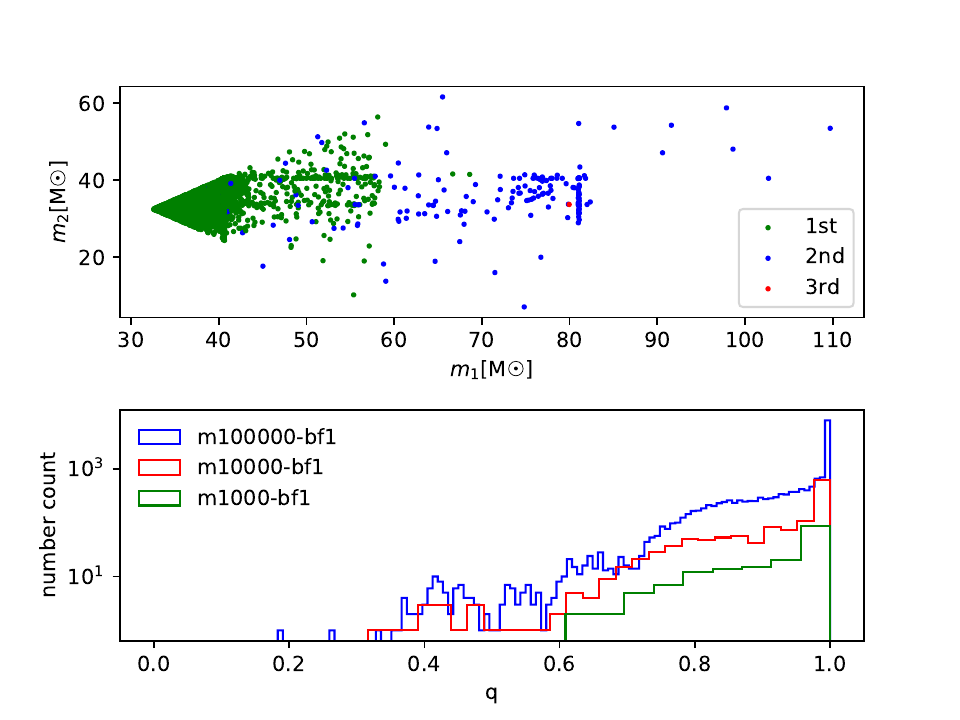}
    \caption{Upper panel: $m_1$ versus $m_2$ for each BBH merger from all simulations with primordial binaries. Points of different colors indicate the generation of the merger. 
    Lower panel: histogram of the mass ratio (q). }
    \label{fig:q_bf1}
\end{figure}

\begin{figure}
	\includegraphics[width=\columnwidth]{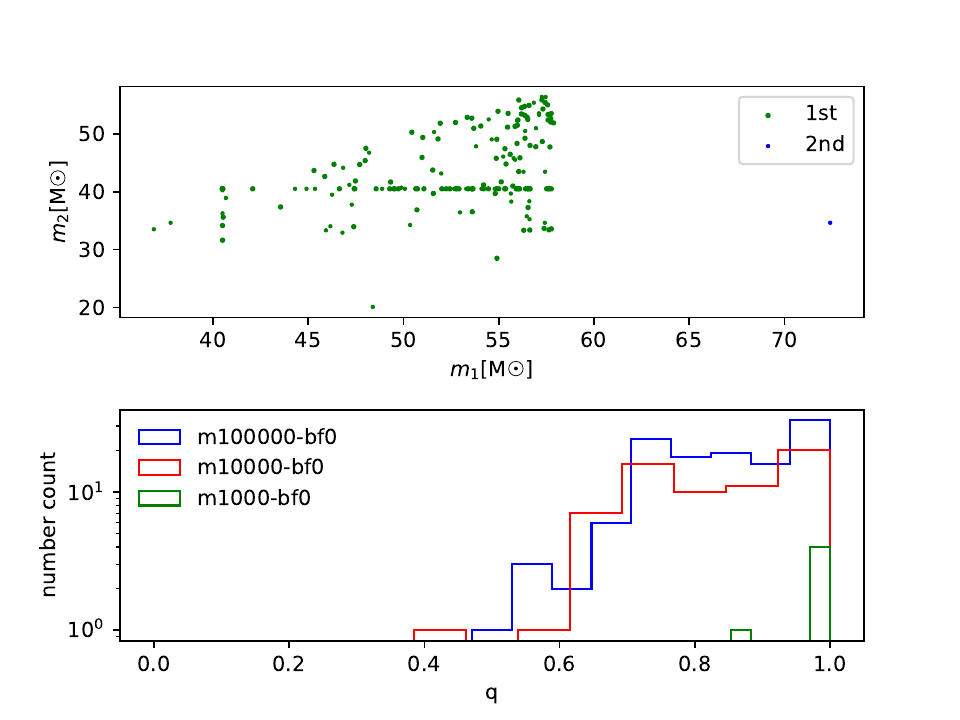}
    \caption{Similar to Figure~\ref{fig:q_bf1}, but showing results from simulations without primordial binaries.
    }
    \label{fig:q_bf0}
\end{figure}

To calculate $v_{\mathrm{k}}$, we need to obtain the $q$ and spin parameters of BHs.
Figure~\ref{fig:q_bf1} and Figure~\ref{fig:q_bf0} illustrate the distribution of $m_1$, $m_2$, and histogram of $q$ for BBHs forming PIBHs in clusters with and without primordial binaries, respectively. 
Most PIBHs are formed through a single merger, with a few undergoing second-generation mergers, and only one example formed through a third-generation merger.
The values of $q$ are predominantly distributed between 0.5 and 1. With primordial binaries, the merging BBHs tend to have comparable masses with a peak of $q$ near 1. And in star clusters with initial masses to be $10000M_\odot$ and $100000M_\odot$, there are also a few cases where the mass ratio is less than 0.5.

\begin{figure}
	\includegraphics[width=\columnwidth]{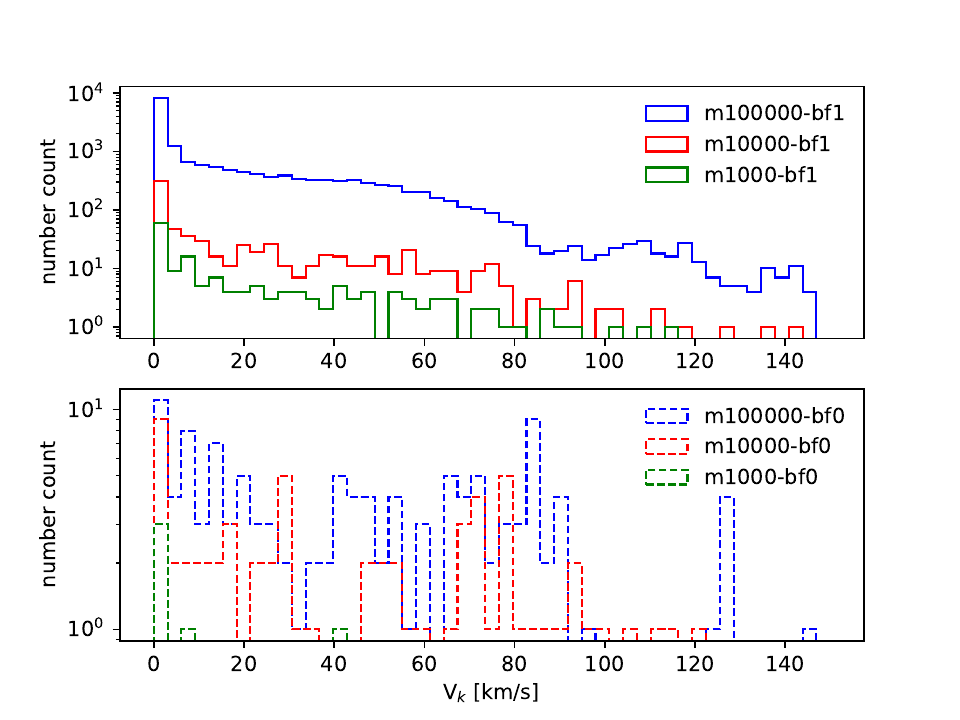}
    \caption{ Considering only the impact of the BBH mass ratio on the GW recoil, the kick velocity of the BH produced during the BBH merger with primordial binaries. 
    }
    \label{fig:vkick_bf1}
\end{figure}

We first analyze the distribution of \( v_{\mathrm{k}} \) for BBH mergers that generate PIBHs, excluding the spins of individual BH components. 
This provides a lower limit for \( v_{\mathrm{k}} \). Figure \ref{fig:vkick_bf1} shows the distribution of \( v_{\mathrm{m}} \) without spins for models with and without primordial binaries, respectively. 
Most of \( v_{\mathrm{k}} \) are below 20 km/s, with the maximum value not exceeding 150 km/s. This indicates that, for most cases, the post-merger PIBHs experience relatively low recoil, which may not be sufficient to eject them from their original clusters.

We now examine the influence of spins on \( v_{\mathrm{k}} \). Including BH spins is critical, as spins introduce complexities in the recoil behavior by interacting with the orbital angular momentum, potentially amplifying or counteracting the momentum transfer to the merged BH. 

\begin{figure}
	\includegraphics[width=\columnwidth]{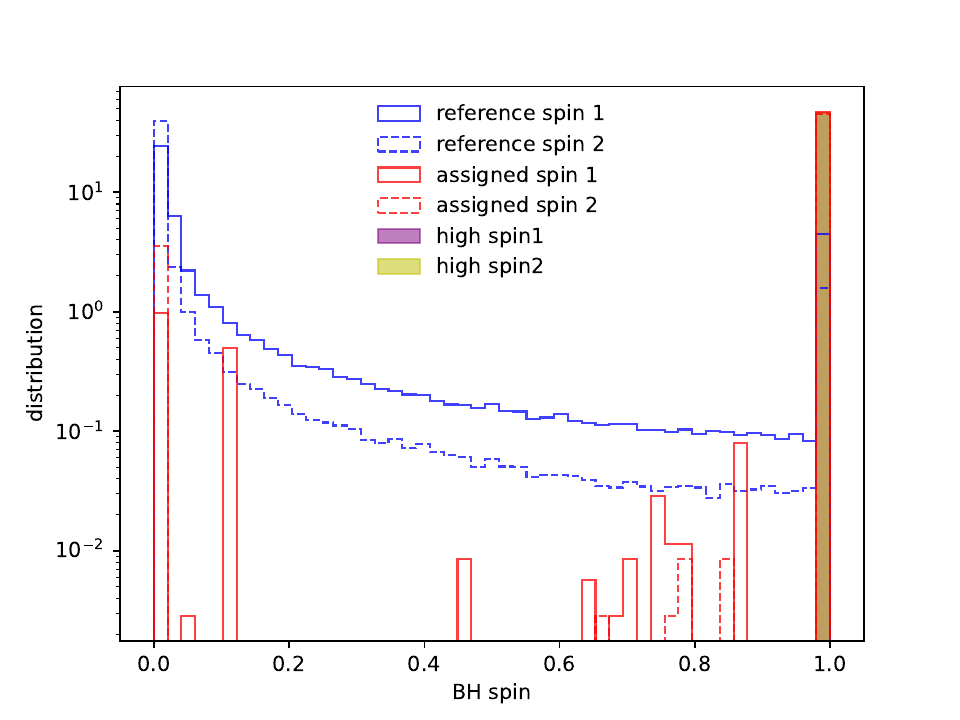}
    \caption{The spin distribution of BH components in binary systems within Pop III star clusters (red) and isolated binary evolution reference models (blue). Solid and dashed lines represent the spin distributions of the primary and secondary components, respectively. Additionally, the shaded histogram in purple and green highlights the high spins of BHs in BBHs that have involved tidally spin-up. 
    }
    \label{fig:spin distri}
\end{figure}

Using the method described in Section~\ref{sec:spin}, we assign spins to BBH components in Pop III star clusters based on the isolated binary evolution model and compare their spin distribution differences, as shown in Figure~\ref{fig:spin distri}.

The results show two distinct populations of BH spins: a low-spin group and a high-spin group.
The high-spin group corresponds to BBHs originating from interacting binaries, where the component spins are enhanced through tidally spin-up during post-common envelope phase because of their small separations \citep{2016MNRAS.462..844K, 2017ApJ...842..111H}. In contrast, the low-spin group represents BBHs formed from other channels, exhibiting a broad distribution with a peak near zero spin.

Due to stellar dynamics, the orbital parameters of BBHs in Pop III star clusters differ from those of the reference BBH populations formed through isolated binary evolution. 
Consequently, the BH spins in Pop III star clusters also deviate from those in isolated binary systems, as shown in Figure~\ref{fig:spin distri}. 
BHs in Pop III clusters exhibit a more extreme distribution, with spins concentrated near zero or unity, whereas isolated binary evolution produces a broader range of spins.

\begin{figure}
    \includegraphics[width=\columnwidth]{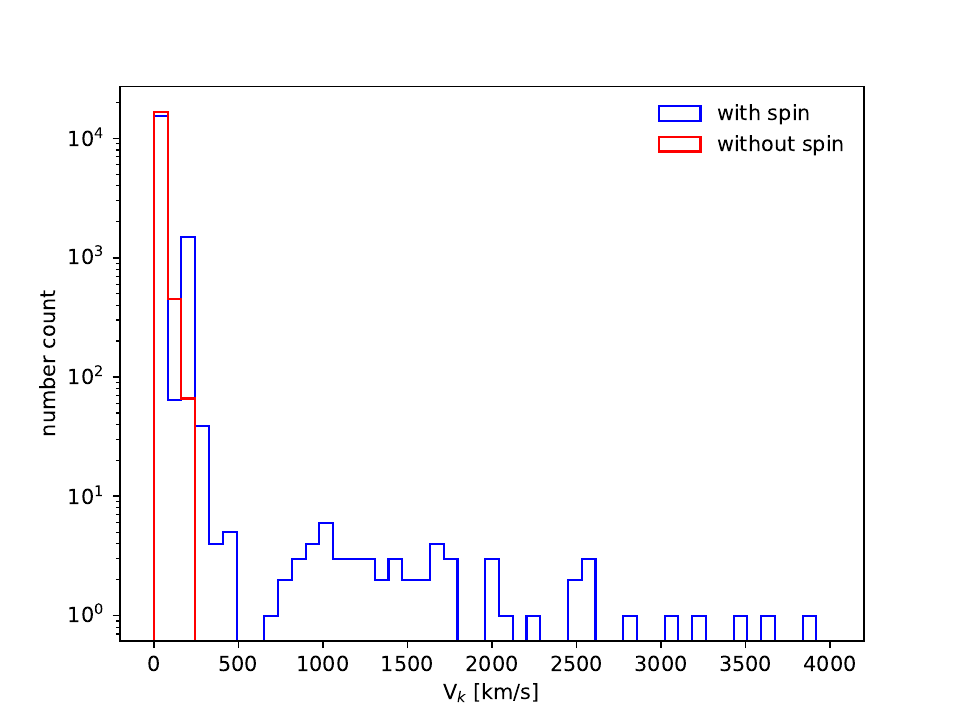}
    \caption{ Considering with and without the spins of the BBHs, the kick velocity of the BH produced during the BBH merger for the m100000-bf1 model. 
    }
    \label{fig:vkick_spin}
\end{figure}

Using these spin values, we randomly sample the spin orbit angle, recalculate \( v_{\mathrm{k}} \), and show its histogram in Figure~\ref{fig:vkick_spin}.
The \( v_{\mathrm{k}} \) without spin is shown for comparison. 
The inclusion of spin significantly broadens the \( v_{\mathrm{k}} \) distribution, particularly introducing a secondary peak around 200 km/s and extending the distribution up to 4000 km/s.
indicating that spin plays a critical role in amplifying the recoil velocity after BBH mergers.

Due to recoil, some PIBHs will immediately escape from clusters that cannot produce secondary mergers. To estimate the fraction of these escaped PIBHs, we need to calculate the escape velocities of Pop III star clusters. A higher escape velocity increases the likelihood that a BH will remain within the cluster.

\begin{figure}
	\includegraphics[width=\columnwidth]{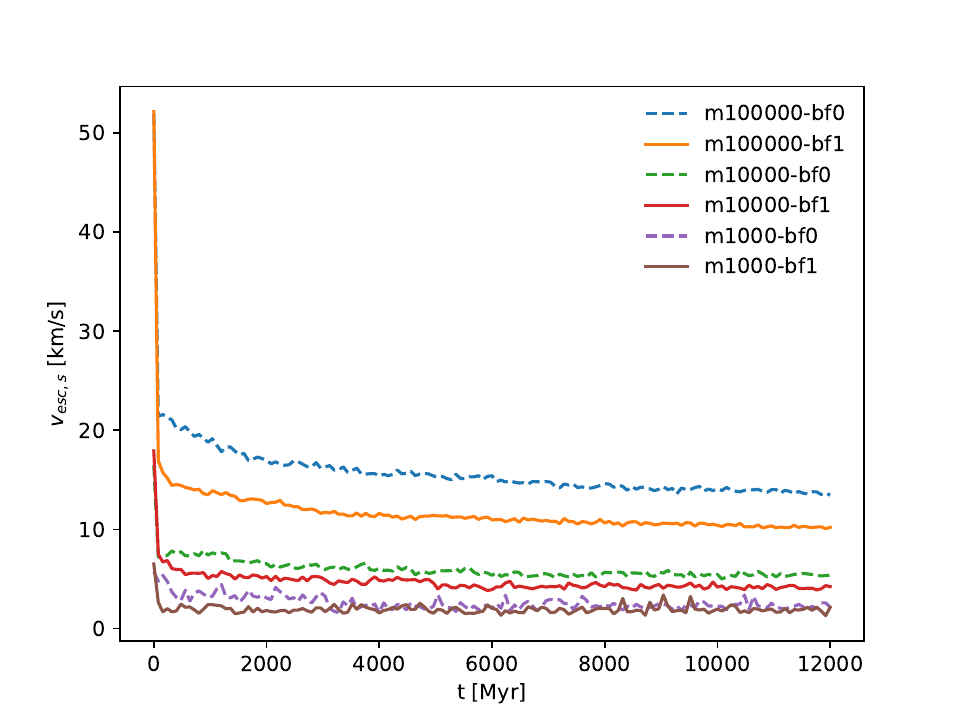}
    \caption{ Evolution of the escape velocity only from stars of the star cluster over time. 
    }
    \label{fig:esc}
\end{figure}

For Pop III star clusters, both stellar mass and the dark matter halo contribute to the cluster's gravitational potential and determine the escape velocity. We estimated the escape velocities from these two components separately.

The escape velocity derived solely from the stellar component ($v_{\text{esc,s}}$) is approximated by using the central escape velocity of the King model.
The central potential \( \Phi(0) \) for the King model can be approximated as:
\begin{equation}
\Phi(0) \approx -\frac{G M}{r_\mathrm{h}} \left( \frac{W_0}{3} \right)
\end{equation}
where \( G \) is the gravitational constant, \( M \) is the total mass of the cluster, and \( r_\mathrm{h} \) is the half-mass radius of the cluster.
the central escape velocity \( v_e \) for a King model with \( W_0 = 9 \) is given by:
\begin{equation}
v_{\text{esc,s}} =\sqrt{2 |\Phi(0)|} = \sqrt{\frac{6GM}{r_\mathrm{h}}}.
\end{equation}
As illustrated in Figure~\ref{fig:esc}, $v_{\text{esc,s}}$ rapidly declines within a few Myr due to the strong stellar wind mass loss from massive stars, which reduces the gravitational potential of the cluster. 
Subsequently, $v_{\text{esc}}$ stabilizes. 
Clusters without primordial binaries show a higher $v_{\text{esc,s}}$ than those with binaries, indicating that primordial binaries contribute to additional heating and drive stronger expansion of the cluster.

The additional central escape velocity provided by the dark matter halo, as discussed in \cite{Wang2022}, is given by:
\begin{equation}
v_{\text{esc,h}} = \sqrt{\Psi(\infty) - \Psi(0)} = \sqrt{\frac{G M_{\text{vir}}}{{r_s} \left[ \log \left(1 + C \right) - \frac{C}{1 + C} \right]}}
\end{equation}
For our simulations, where the halo concentration parameter C is set to 15.3, the central escape velocity attributable to the halo component is approximately 53 km/s.

Considering both stellar components and the dark matter halo, we can determine the cluster's overall escape velocity ($v_{\text{esc}} = v_{\rm esc,s } + v_{\rm esc,h}$), and the ejected fraction of PIBHs formed from BBH mergers due to GW kicks ($f_{\mathrm{e}}$). 
For example, in model m100000-bf1, PIBHs with $v_{\mathrm{k}}$ over 65 km/s. 
Assuming no initial spins, $f_{\mathrm{e}}$ is 3.7\%.
However, when accounting for BH spin, this fraction increased to $9.9\%$. Table~\ref{tab:number} lists the values of $v_{\text{esc}}$ and the ejected fraction of PIBHs $f_{\mathrm{e}}$ for various models.

In the m100000-bf1 model, 59\% of PIBH-BH mergers are excluded due to PIBH ejection from the cluster, and this represents a lower limit.
These mergers are omitted from the counts for different models in Table~\ref{tab:number} and subsequent analysis.

\subsubsection{Properties of BBH mergers involving PIBHs}

Table~\ref{tab:number} shows PIBH-BH mergers per cluster across different cluster masses \(M\) and primordial binary fractions. 
Excluding escaped PIBHs from GW kicks, PIBH-BH mergers formed via BBH mergers are about 6 times more frequent than those from the stellar collision channel in the m100000-bf1 model.
Clusters without primordial binaries produce far fewer PIBH-BH mergers, highlighting the crucial role of primordial binaries. 
PIBH formation through stellar collisions is unlikely without primordial binaries.

The PIBH-BH merger counts scale roughly linearly with $M$, similar to PIBH formation via BBH mergers.
For example, in m100000-bf1, each cluster contains average 0.548 mergers, while lower-mass clusters like m10000-bf1 and m1000-bf1 show decreases by factors of 10 and 100, respectively. 
In addition, PIBH-BH mergers account for 0-0.1 of total BBH mergers and show no clear dependence on \(M\) or the presence of primordial binaries.

\begin{figure}
	\includegraphics[width=\columnwidth]{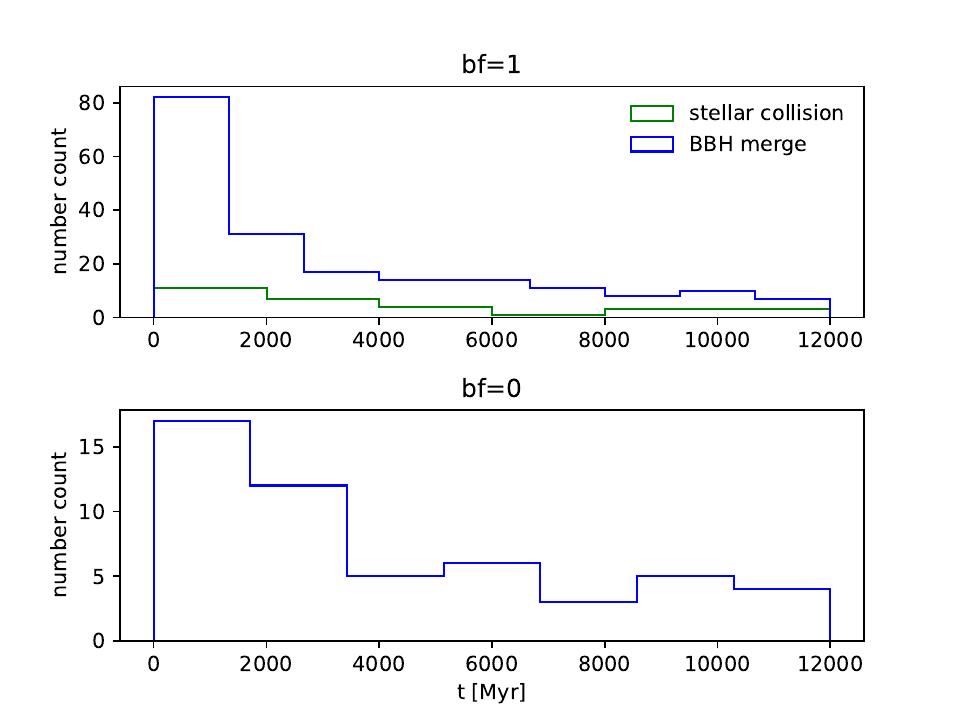}
    \caption{ The time distribution of PIBH-BH mergers for m100000 models. 
    }
    \label{fig:bh_merge}
\end{figure}

Figure~\ref{fig:bh_merge} illustrates the time distribution of PIBH-BH mergers for the m100000 models, 
comparing clusters with and without primordial binaries. 
The number of mergers in m10000 and m1000 models is insufficient for reliable statistical analysis and is therefore excluded from the plot.
Notably, a significantly higher number of BBH mergers occur within the first  \( 2000 \) Myr in the cluster evolution, with most of the merging BBHs containing PIBHs originating from the BBH merger channel.

\begin{figure}
	\includegraphics[width=\columnwidth]{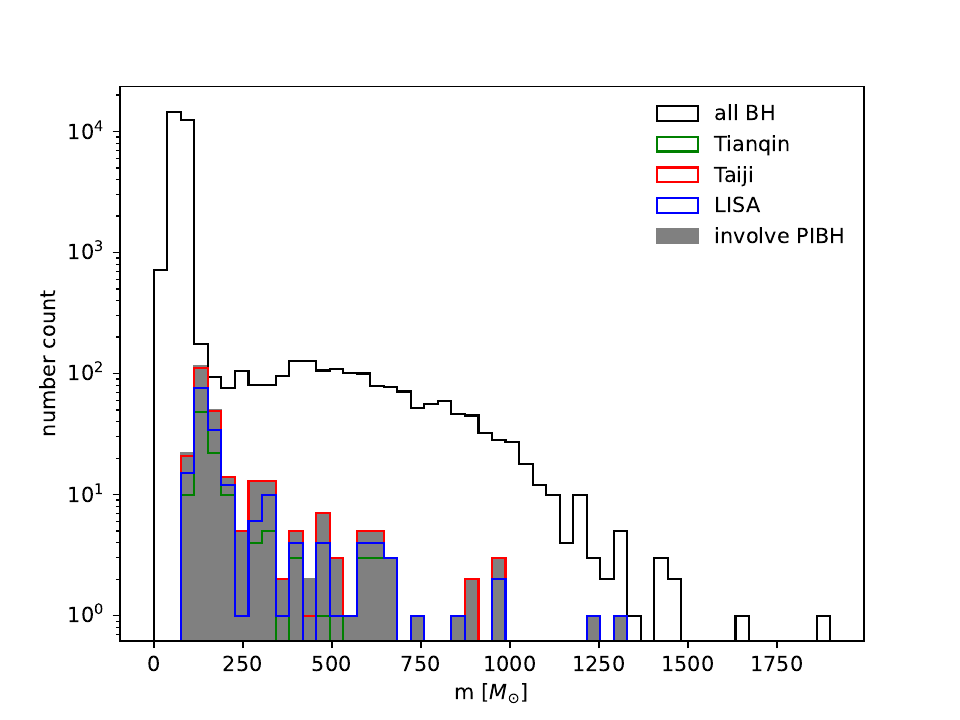}
    \caption{The BH mass spectrum from all merger events across all models, taking into account the effects of GW recoil. The shaded area represents the spectrum from PIBH-BH merger, with colored histograms representing detectable events by three space-borne detectors.
    }
    \label{fig:allbhmass}
\end{figure}

Figure~\ref{fig:allbhmass} shows the BH mass spectrum from all merger events, dominated by masses below 200 $M_\odot$, with numbers decreasing as mass increases, reaching over 1800 $M_\odot$. 
The spectrum for PIBH-BH mergers is also shown, along with detectable events for three space-borne GW detectors. 
PIBH-BH mergers include both PIBH with lower-mass BHs and a significant population of IMBH-PIBH mergers. 
Space-borne detectors can cover the entire mass range.

\subsection{Peak frequency and characteristic strain for GW detection}

The peak frequency ($f_{\rm peak}$) of GWs emitted during the merger of BBHs (BBHs) is crucial as it represents the frequency with the maximum radiation power. It can be calculated using a formula that considers the masses of the BBHs, their orbital eccentricity e, and their semi-major axis a \citep{2021RNAAS...5..275H}:

\begin{equation}
    \begin{aligned}
f_{\text{peak}} &= \frac{\sqrt{G (m_1 + m_2)}}{\pi}  \\
&\quad \times \left( 1 - 1.01678 \times 10^{5} - 5.57372 \times 10^{2} \right. \\
&\qquad\left. - 4.9271 \times 10^{3} + 1.68506 \times 10^{4} \right) \\
&\quad \times \frac{1}{\left[a \left(1 - e^2\right)\right]^{1.5}}.
\end{aligned}
\end{equation}

As BBHs evolve, their orbital eccentricity decreases due to GW radiation, leading to a transition from eccentric to more circular orbits. The evolution curves are given in \cite{peters1964gravitational}:
\begin{equation}
\dot{a} = -\frac{\beta F(e)}{a^3},
\end{equation}

\begin{equation}
\dot{e} = -\frac{19}{12} \frac{\beta e}{a^4 (1 - e^2)^{5/2}} \left( 1 + \frac{121}{304} e^2 \right),
\end{equation}

\begin{equation}
F(e) = \sum_{n=1}^{\infty} g(n, e) = \frac{1}{(1 - e^2)^{7/2}} \left( 1 + \frac{73}{24} e^2 + \frac{37}{96} e^4 \right),
\end{equation}

\begin{equation}
\beta = \frac{64}{5} \frac{G^3 m_1 m_2(m_1 + m_2)}{ c^5}.
\end{equation}
Here \( g(n, e) \) is described by equation(20) from \cite{peters1963gravitational}. And the evolution continues until \( a \) approach to zero.

It is influenced by the frequency of the waves, the masses of the merging BBH, their distance from the observer, and the orbital eccentricity. The emission of GWs includes different order harmonics, the correspounding characteristic strain \( (h_{c,n}) \) of the \( n \)th order harmonic of inspiral GW is given by \citep{kremer2019post}: 
 
\begin{equation}
h_{c,n}^{2} = \frac{2}{3\pi^{4/3}} \frac{G^{5/3}}{c^3} \frac{M_{c,z}^{5/3}}{D_{L}^2} \frac{1}{f_{n,z}^{1/3}(1 + z)^2} \cdot \frac{2^{2/3}}{n} \frac{g(n, e)}{F(e)}
\label{eq:h_c,n}
\end{equation}

where \( M_{c,z} \) is the observed chirp mass at redshift \( z \), $D_{L}$ is the luminosity distance, 
and \( f_{n,z} \) is the observed frequency of the \( n \)th harmonic. The Equation\eqref{eq:h_c,n} applies to describing the characteristics of the inspiral phase, but as the BHs enter the merger and ringdown phases, the orbital eccentricity of BBHs gradually decreases to nearly zero due to the dissipative effect of GW radiation. At this stage, the PhenomD waveform is used to describe the GW characteristics of the subsequent phases \citep{husa2016frequency,khan2016frequency}. 

\begin{figure}
	\includegraphics[width=\columnwidth]{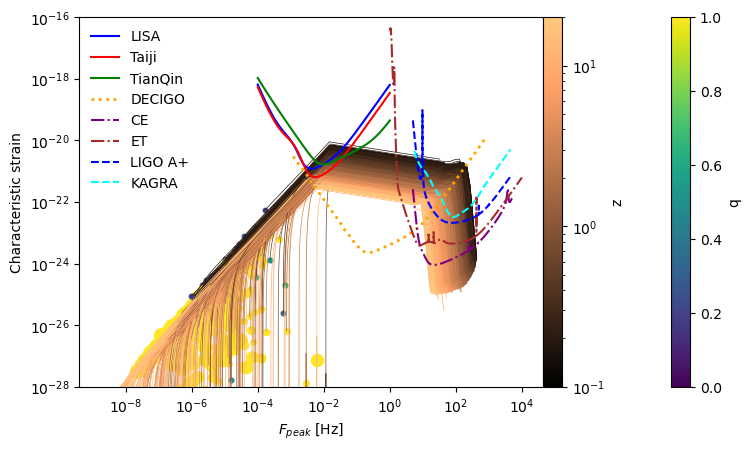}
    \includegraphics[width=\columnwidth]{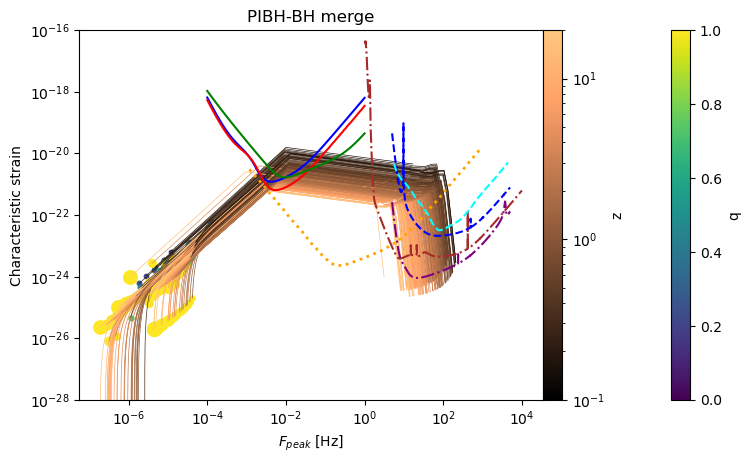}
    \caption{ The characteristic strain of GW (GW) signals as a function of the peak frequency. The upper panel is for all BBH merger pathways leading to the formation of PIBHs across all the simulations, and the lower panel is for all PIBH-BH mergers across all the simulations. The color of the terminal dots represents the mass ratios of the merging binaries, while the color gradient of the lines indicates the redshift of these events. }
    \label{fig:strain}
\end{figure}

Figure~\ref{fig:strain} illustrates the characteristic strain of GW signals as a function of their peak frequency, specifically focusing on mergers involving PIBHs. The analysis includes information about the redshift (z) and mass ratio of each merger event, which are represented in the color gradient of the plotted lines and the terminal dot. The upper panel displays the GW signals produced through the BBH merger channel that lead to the formation of PIBHs, based on simulations of the six models. The lower panel shows the same relationship but for all PIBH-BH mergers across all the simulations, excluding the events that include PIBHs that can be ejected by GW kicks.
The sensitivity curves of various GW detectors are also plotted, including space-borne detectors like LISA, Taiji, TianQin, and DECIGO, as well as ground-based detectors such as advanced LIGO (aLIGO) and KAGRA. At the same frequency, if the value of the characteristic strain exceeds the sensitivity curve of the detector, it indicates that the signal has the potential to be detected. Some low-redshift merger events can be detected by almost all detectors. Specifically, the detection rates for Tianqin, Taiji, and LISA are estimated as  43.4\%, 97.8\%, and 66.4\%, respectively. DECIGO, CE, and ET, however, have a broader detection capability, allowing them to detect nearly all merger events related to PIBHs.

\subsection{PIBH-BH merger rate during cosmic evolution}

In this section, we analyze the PIBH-BH merger rates during cosmic evolution. Table~\ref{tab:rate} presents the 
estimated rates based on the stellar mass density of Pop III stars, using upper and average star formation rate limits derived from cosmological data. 
The average star formation rate, from the numerical models of \cite{skinner2020cradles}, corresponds to a total stellar mass of approximatedly \(3.2 \times 10^4 \, M_{\odot} \, \text{Mpc}^{-3}\) over the redshift range from 20 to 10. The upper limit, from \cite{inayoshi2021gravitational}, is \(2 \times 10^5 \, M_{\odot} \, \text{Mpc}^{-3}\). 
Using these values, PIBH-BH merger rates can be estimated for each simulation model with a given cluster mass $M$, assuming all Pop III star forms in clusters of that mass.
While this assumption is unrealistic, it provides insight into how merger rates depends on initial cluster mass. 

As the mass of the cluster increases (from M1000 to M100000), the PIBH-BH merger rate does not show a simple linear trend. 
The m10000-bf1 and m100000-bf1 models show comparable rates, while m10000-bf0 has a slightly higher rate than m100000-bf0. 
For the lowest-mass clusters, m1000-bf1 shows a lower but similar order of magnitude in rates, whereas m1000-bf0 shows no PIBH-BH mergers. 
This absence may be due to stochastic effects, as the total simulation mass for m1000 is much smaller than that of the other two \(M\) models. 
Overall, models with primordial binaries exhibit significantly higher PIBH-BH merger rates, indicating that merger rates are weakly dependent on \(M\) but strongly influenced by the presence of primordial binaries.

\begin{table*}
	\centering
    \caption{The PIBH-BH merger rates for each model are presented, including the average, upper limit based on the total stellar mass of Pop III stars, and the rate per unit mass. 
    Rates from young star clusters in previous studies are shown for comparison. }

	\label{tab:rate}
	\begin{tabular}{lcccccc} 

        \hline
		Model name & m100000-bf1 & m100000-bf0 & m10000-bf1 & m10000-bf0 & m1000-bf1 & m1000-bf0 \\
  		\hline 
		Average [$\text{yr}^{-1} \text{Gpc}^{-3}$] & 0.017 & 0.005 & 0.016 & 0.008 & 0.010 & 0.000 \\
        
        Uplimit [$\text{yr}^{-1} \text{Gpc}^{-3}$] & 0.106 & 0.030 & 0.100 & 0.050 & 0.060 & 0.000 \\
        
        Per mass [$M_{\odot}^{-1}$] & $5.5 \times 10^{-6}$ & $1.8 \times 10^{-6}$ & $4.3 \times 10^{-6}$ & $3.0 \times 10^{-6}$ & $3.3 \times 10^{-6}$ & $0.0$ \\
	\end{tabular}
    
        \begin{tabular}{lcccc}
        \hline
         Young star clusters & \cite{DiCarlo2020a} & \multicolumn{2}{c}{\cite{Kremer2020}} & \cite{Banerjee2022}\\ 
        \hline
        Mass & $10^3-3\times10^4 M_\odot$ &   \multicolumn{2}{c}{$4\times10^5 M_\odot$} & $7.5\times 10^4 M_\odot$\\ 
        Rate & $ < 2.5 \times 10^{-7}(M_{\odot}^{-1})$ & $2 \times 10^{-6}(M_{\odot}^{-1})$ & $3 \times 10^{-5}(M_{\odot}^{-1})$ & $0 - 0.84 (\text{yr}^{-1} \, \text{Gpc}^{-3})$ \\
        Channel
         & all & stellar collision & BBH merger &  all \\ 
		\hline
	\end{tabular}
\end{table*}

We also compare our results for PIBH-BH merger rates in Pop III clusters with those formed in young massive clusters, globular clusters and nuclear star clusters reported in previous studies, as shown in Table~\ref{tab:rate}
In studies of young star clusters with mass range of $10^3 - 3\times10^4 M_\odot$ with metallicity $Z=0.002$ from \cite{DiCarlo2020b,DiCarlo2021}, the PIBH-BH merger rate is \(<2.5 \times 10^{-7} \ M_{\odot}^{-1}\). \cite{Kremer2020} studied more massive young star clusters and suggests that PIBH-BH is higher, with the value about  \(2 \times 10^{-6} \ M_{\odot}^{-1}\) through stellar evolution and \(3 \times 10^{-5} \ M_{\odot}^{-1}\) through BBH megers.
\cite{Banerjee2022} find that, in a YSCs of initial mass \(7.5 \times 10^4 \, M_{\odot}\), metallicity \(0.0002 \leq Z \leq 0.02\), the overall merger rate is between 0 and 0.84 \(\text{yr}^{-1} \, \text{Gpc}^{-3}\). 
\cite{Mapelli2021} compared the fraction of PIBH-BH mergers in nuclear star clusters, globular clusters, and young star clusters, finding values of 0.05, 0.02, and 0.007, respectively, for $Z=0.002$. This suggests that denser nuclear and globular clusters may produce more PIBH-BH mergers per unit mass. 
At solar metallicity, all fractions decrease by at least an order of magnitude, indicating a preference for low-metallicity environments.
The PIBH-BH merger rate per unit mass in Pop III star clusters with primordial binaries is lower than that in young massive star clusters, but higher than that in young star clusters, highlighting the significance of Pop III star clusters as a potential environment for producing GW190521-like events.

\section{Discussion}
\label{sec:discussion}

As described in section \ref{sec:Initial_condition}, isolated binaries cannot form PIBH-BH mergers in the L model. However, dynamical interactions generate PIBH-BH mergers quite efficiently even in the L model. The average merger rate is $\sim 0.014$ yr$^{-1}$ Gpc$^{-3}$ when the binary fraction is set to unity. This average merger rate is comparable to the PIBH-BH merger rate formed through isolated binary evolution in the M model \citep{Tanikawa2021}, or the merger rate of GW190521-like events \citep{Abbott2020,Abbott2020b}. Pop III star clusters should be a promising formation site of PIBH-BH mergers regardless of Pop III star evolution models.

To analyze the plausibility of the initial mass of Pop III star clusters used in our simulations, we can look at theoretical predictions from both cosmological hydrodynamic and magnetohydrodynamic simulations.
According to cosmological hydrodynamic simulations by \cite{Sakurai2017} , Pop III star clusters are likely to form within dark matter halos with masses around \(10^7 \, M_{\odot}\), typically forming at redshifts around \(z \sim 20\). These clusters generally have masses up to \(10^5 \, M_{\odot}\), with a distribution leaning toward clusters below this threshold. Additionally, \cite{Wang2022} corroborates this estimate, suggesting that Pop III clusters tend to have masses \(M_{\text{cl}} \lesssim 10^5 \, M_{\odot}\).

Magnetohydrodynamic simulations by \cite{he2019simulating} further support these findings, showing that molecular clouds in environments resembling high-redshift galaxies can produce star clusters up to about \(10^5 \, M_{\odot}\), with the majority of clusters clustering around \(10^4 \, M_{\odot}\). This suggests that star clusters with initial masses around \(10^4 \, M_{\odot}\) are likely more common, while more massive clusters approaching \(10^5 \, M_{\odot}\) might be rarer.

Given this context, the initial cluster masses of $1000 \, M_{\odot}$, $10,000 \, M_{\odot}$, and $100,000 \, M_{\odot}$ used in our simulations span a reasonable range in comparison with the theoretical predictions. Clusters with initial masses of $1000 \, M_{\odot}$ and $10,000 \, M_{\odot}$ are well within the expected mass range based on these simulations and align with predictions for the common formation pathways of Pop III star clusters. The largest clusters in our simulations, with initial masses of $100,000 \, M_{\odot}$, represent the upper limit of plausible cluster masses and may account for rarer formation events. These clusters provide insights into the merger rate and formation efficiency in the high-mass regime, complementing the typical cluster sizes modeled by \cite{Sakurai2017} and \cite{he2019simulating}.

PIBH can also form via other scenarios that are not the focus of this study.
The primordial BHs formed through gravitational instability collapse in the early universe might form BHs in a wide mass region, including PIBHs \citep{Carr2016,Carr2020,DeLuca2021,Clesse2022}.
In the accretion disks of active galactic nuclei (AGNs) or gas-rich nuclear star clusters, low-mass BHs may grow up to a PIBH via accretion and dynamically form BBHs\citep{Mckernan2012, Samsing2022,Tagawa2021,Tagawa2021b,WangJM2021,Vajpeyi2022,Natarajan2021}.

The GW190521 event may also be explained as a merger without PIBHs depending on how to interpret the event signals.
\cite{Nitz2021} and \cite{Cui2023} suggest that the GW190521 can also be explained by an intermediate-mass ratio inspiral with two component masses of $166 M_\odot$ and $16 M_\odot$.
\cite{Chen2019} and \cite{Zhang2023} pointed out that if the event occurs close to a super massive BH, a low-mass BBH merger at a redshift of $0.01-0.1$ with components mass of $10-20 M_\odot$ could masquerade as a BBH with PIBH at redshift around 1.

Due to the limited information from the GW190521 event, we still cannot confirm its formation scenario. But future GW detectors, including the Einstein telescope \citep{Santoliquido2024,punturo2010einstein,maggiore2020science} and Cosmic Explorer \citep{reitze2019cosmic}, and future spaceborne detectors, such as LISA \citep{amaro2017laser} Taiji and TianQin \citep{Liu2022} are expected to catch Pop III BBH merger events \citep{sesana2009lisa}, the measurement of eccentricity of GW events can be significantly improved, which is an important parameter to distinguish different formation channels \citep{Romero-Shaw2020}.
Therefore, studying various scenarios in detail is essential for better predicting event properties, which is another aim of this study regarding PIBHs originating from Pop III star clusters.

\section{conclusion}
\label{sec:conclusion}

This study explores the formation and evolution of BHs in the pair-instability mass gap (65--120 $M_\odot$) in Pop III star clusters. Using advanced $N$-body simulations with initial cluster masses of 1,000, 10,000, and 100,000 $M_\odot$ and varying primordial binary fractions, we investigated the primary formation channels of these PIBHs: stellar collisions and BBH mergers and their subsequent GW signals.

BBH mergers are the dominant formation channel for PIBHs, contributing significantly more than stellar collisions. For example, in the M100000-bf1, 49.09 PIBHs are formed on average per cluster via BBH mergers, compared to only 1.89 via stellar collisions. The number of PIBHs decreases linearly by 10 for every 10-unit reduction in $M$.

Continued mergers of PIBHs are possible, particularly in environments with high escape velocities. GW recoils play a critical role in determining whether PIBHs remain in clusters or are ejected. In dense clusters, high escape velocities increase the preservation of BHs. In the model M100000-bf1, 10\% of BBH mergers producing PIBHs result in ejections from the cluster due to GW recoils when spins are considered. 

The PIBH-BH merger rate is weakly dependent on host cluster mass but strongly influenced by the presence of binaries. With primordial binaries, the average rate is 0.010-0.017 \(\text{yr}^{-1} \text{Gpc}^{-3}\) with an upper limit of 0.060-0.106 \(\text{yr}^{-1} \text{Gpc}^{-3}\). Without primordial binaries, the average rate drops to 0-0.008  \(\text{yr}^{-1} \text{Gpc}^{-3}\), with a maximum of 0-0.05 \(\text{yr}^{-1} \text{Gpc}^{-3}\). 
These rates are comparable to those in youmg massive star clusters, highlighting Pop III star clusters as significant environments for producing GW190521 like events.

Analyzing the characteristic strain of GW events shows that 40-80\% of PIBH-BH mergers can be detected by future space-borne detectors like LISA, TianQin, and Taiji, with most covered by DECIGO, ET, and CE. 
As Pop III star clusters form above redshift 10, they may significantly contribute to high-redshift PIBH-BH events.

\section{Acknowledgements}

L.W. thanks the support from the National Natural Science Foundation of China through grant 21BAA00619 and 12233013,  the one-hundred-talent project of Sun Yat-sen University, the Fundamental Research Funds for the Central Universities, Sun Yat-sen University (22hytd09). Shuai Liu thanks the support from Zhaoqing City Science and Technology Innovation Guidance Project (No. 241216104168995) and the Young Faculty Research Funding Project of Zhaoqing University (No. qn202518). This work was supported in part by JSPS KAKENHI Grant Numbers 23K22530 and 22H01259.

\software{numpy (\citealp{harris2020array}),
          matplotlib (\citealp{Hunter:2007}),
          \textsc{petar} (\citealp{Wang2020b,PeTar_zenode}, https://github.com/lwang-astro/PeTar),
          \textsc{sdar} (\citealp{Wang2020a}, https://github.com/lwang-astro/SDAR),
          \textsc{bseemp} (\citealp{Tanikawa2020}, https://github.com/atrtnkw/bseemp),
          \textsc{mcluster} (\citealp{kupper2011mass}, https://github.com/lwang-astro/mcluster)
        }

\bibliography{paper}{}
\bibliographystyle{aasjournal}

\end{CJK*}
\end{document}